\documentclass[%
onecolumn,
amsmath,amssymb,
aps,
]{revtex4-2}

\usepackage{graphicx}% Include figure files
\usepackage{dcolumn}% Align table columns on decimal point
\usepackage{bm}% bold math
\usepackage{subcaption}
\usepackage{ragged2e}
\usepackage{comment}
\renewcommand{\selectlanguage}[1]{} % Need this to stop .bbl file from complaining about having language = {en} entries in the bib

\usepackage{xcolor}
\definecolor{OliveGreen}{rgb}{0.1, 0.4, 0.1}

\usepackage{soul}
\setstcolor{red}

\usepackage{setspace}
\emergencystretch=\maxdimen
\allowdisplaybreaks

\usepackage[colorlinks=true,citecolor=OliveGreen,linkcolor=OliveGreen,urlcolor=OliveGreen]{hyperref}

\makeatletter
\newcommand{\fcolon}{%
  \mathrel{\mathpalette\fcolon@\relax}%
}
\newcommand{\fcolon@}[2]{%
  \sbox\z@{$\m@th#1:$}%
  \vbox to\ht\z@{%
    \hbox{$\m@th#1.$}%
    \vss
    \vss
    \hbox{$\m@th#1.$}%
    \vss
    \hbox{$\m@th#1.$}%
  }%
}
\makeatother

\begin{document}
\title{Kinetic Theory of Binary Fluid–Surfactant Systems: A Variational Framework}
\author{Alexandra J.\ Hardy}
\author{Samuel Cameron}
\author{Steven McDonald}
\author{Abdallah Daddi-Moussa-Ider}
\author{Elsen Tjhung}
\email{elsen.tjhung@open.ac.uk}
\thanks{corresponding author}
\affiliation{School of Mathematics and Statistics, The Open University, Walton Hall, Kents Hill, Milton Keynes, MK7 6AA, United Kingdom.} 
\date{\today}

\begin{abstract}
We derive a self-consistent hydrodynamic theory of coupled binary-fluid–surfactant systems from the underlying microscopic physics using Rayleigh’s variational principle. 
At the microscopic level, surfactant molecules are modelled as dumbbells that exert forces and torques on the fluid and interface while undergoing Brownian motion. 
We obtain the overdamped stochastic dynamics of these particles from a Rayleighian dissipation functional, which we then coarse-grain to derive a set of continuum equations governing the surfactant concentration, orientation, and the fluid density and velocity. 
This approach introduces a polarization field $\bm{p}(\bm{r},t)$, representing the average orientation of surfactants, which plays a central role in suppressing droplet coalescence.
The remaining hydrodynamic equations are consistently obtained from a mesoscopic free energy functional.
The resulting model accurately captures key surfactant phenomena, including surface tension reduction and droplet stabilization, as confirmed by both perturbation theory and numerical simulations,
and is thermodynamically consistent with both the Gibbs adsorption isotherm and Henry’s law for adsorbed surfactant concentration.
\end{abstract}

\maketitle

\section{Introduction}

\begin{figure}
    \includegraphics[width=0.8\linewidth]{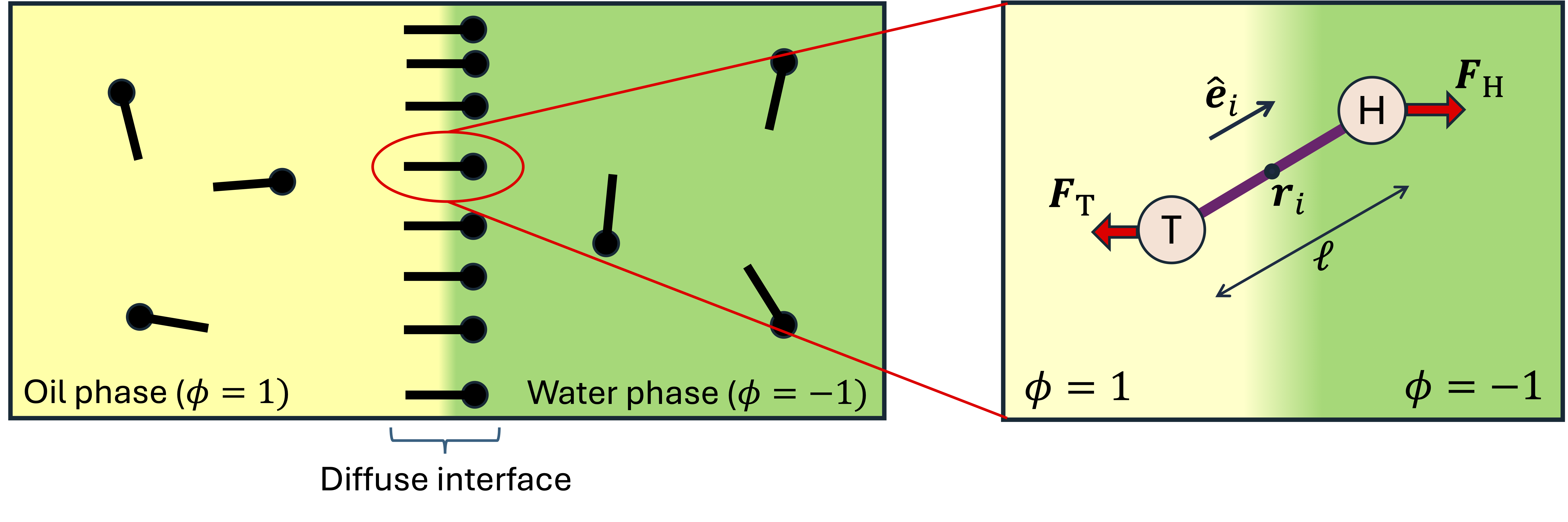}  
    \caption{\justifying
    Left: Schematic diagram illustrating how surfactants (black) are absorbed perpendicularly at the interface between two phases, e.g., water and oil phase (green and yellow respectively). 
    Right: Diagram showing the surfactant molecule modelled as a dumbbell, adsorbed into a diffuse water-oil interface, with `head' H, `tail' T, rod of length $\ell$, centre of mass $\bm{r}_i$, and orientation vector $\hat{\bm{e}}_i$, directed from `tail' to `head'. 
    The fluid exerts a force on each mass point, $\bm{F}_\mathrm{H}$ and $\bm{F}_\mathrm{T}$, due to the hydrophilic/hydrophobic attraction between said mass points and the corresponding fluid phases. 
    The binary fluid order parameter $\phi(\bm{r},t)$ has values between $1$ and $-1$ to represent the two fluid phases.}
    \label{fig:surfactant-diagram}
\end{figure}

Systems containing surfactants are widely studied due to their numerous industrial applications, including in medicine, cleaning products, and the food industry~\cite{tadros2006applied,shaban2020surfactants}. 
Their primary utility lies in their ability to adsorb at fluid-fluid interfaces, such as liquid-gas or oil-water boundaries, where they reduce surface tension and inhibit droplet coalescence~\cite{mulqueen2002theoretical}.
For instance, in an emulsion of oil droplets dispersed in water, surfactants stabilize the droplet interfaces, thereby slowing the phase separation into two macroscopic oil and water phases.
Surfactant molecules are amphiphilic, typically composed of a hydrophilic ‘head’ and a hydrophobic ‘tail’.
The hydrophilic head is attracted to the water phase, while the hydrophobic tail prefers the oil phase, leading to an orientation that is perpendicular to the interface (as shown in Fig.~\ref{fig:surfactant-diagram}, left).
Upon adsorption at the interface, surfactants reduce the interfacial energy and thereby lower the surface tension. 
Once the interface becomes saturated, no further surfactant molecules can adsorb, causing the surface tension to level off~\cite{wang2017modeling,touhami2001study}.
Excess surfactants in the bulk phase may then self-assemble into micelles—spherical aggregates with hydrophobic tails hidden in the core and hydrophilic heads exposed to the surrounding fluid~\cite{santos2016determination,gangula2010analytical}.
Beyond surface tension reduction, surfactants also suppress droplet coalescence~\cite{krebs2012coalescence,dai2008mechanism}, thereby enhancing the stability and mixing of otherwise immiscible phases.

Previous theoretical treatments of ternary systems—comprising two immiscible fluids and a surfactant population—have included Ising-like lattice models~\cite{alexander1978lattice,ahluwalia1996}, Potts models~\cite{gilhoj1996}, and direct molecular dynamics simulations~\cite{laradji2000md}.
Early continuum models either neglected the polarization field entirely or integrated out its degrees of freedom from the dynamics~\cite{kawasaki_continuum_1990,anisimov_landau_1992}.
The first study to explicitly include the coarse-grained polarization field dynamics~\cite{melenkevitz_phase_1997} neglected both hydrodynamic interactions and thermal fluctuations.
More recent continuum models of surfactant-covered binary fluids employ a free energy functional that depends on the surfactant concentration but omits the explicit dynamics of the polarization field $\bm{p}(\bm{r},t)$.
Instead of incorporating $\bm{p}(\bm{r},t)$, such models introduce additional stabilizing terms to prevent the surfactant from destabilizing the diffuse interface~\cite{zhu2020phase,liu2010phase}.
These terms are often chosen to enforce Langmuir’s adsorption isotherm~\cite{kalam2021surfactant} and the Gibbs adsorption equation~\cite{manikantan2020surfactant}, which respectively govern surfactant uptake at interfaces and the relationship between surface tension and surfactant concentration.
As a result, these models can successfully capture the reduction in surface tension with increasing surfactant concentration~\cite{liu2010phase}.
Coupling the surfactant and order parameter fields to the fluid velocity field further allows these models to reproduce, to some extent, the suppression of droplet coalescence~\cite{liu2010phase,soligo2019coalescence,soligo2019breakage}.

A wide range of numerical methods have been employed to directly simulate the continuum equations of diffuse interface systems, including finite volume~\cite{yamashita2024conservative} and finite difference schemes~\cite{teigen2011diffuse}, as well as more specialized computational approaches~\cite{yang2021novel,booty2010hybrid}.
Such simulations provide insights into the behaviour of surfactant-laden droplets under various conditions, including spinodal decomposition~\cite{kim2006numerical}, shear flows~\cite{teigen2011diffuse}, and wetting dynamics~\cite{ganesan2015simulations}.
Alternative approaches to direct numerical simulation include lattice gas~\cite{PhysRevLett.56.1505} and lattice Boltzmann methods~\cite{higuera_lattice_1989,benzi_lattice_1992}, which were initially developed for binary fluid systems~\cite{rothman_immiscible_1988,chan_critical_1990} and later extended to ternary mixtures, where surfactants act as a third component~\cite{love2003multiphase,theissen1999lattice,kian2021multiphase}.
Lattice Boltzmann methods can be broadly classified as either free-energy-based methods or pseudopotential approaches~\cite{kruger_lattice_2017}.
The pseudopotential approach mimics attractive or repulsive interactions between different species of fluid molecules but does not have a direct connection to microscopic physics.
To date, only pseudopotential-based models have been used to explicitly simulate the dynamics of the polarization field associated with surfactant molecules--aside from Ref.~\cite{melenkevitz_phase_1997}, which neglected hydrodynamics.

In this work, we develop a continuum model of surfactant-laden binary fluids that is rigorously grounded in microscopic physics. 
Our central methodology is based on Rayleigh’s minimum energy dissipation principle, which we use to derive the overdamped dynamics of surfactant molecules interacting with a diffuse interface. 
This variational approach allows us to self-consistently obtain both the stochastic microscopic dynamics and, upon coarse-graining, the macroscopic hydrodynamic equations for the coupled fields: binary fluid volume fraction $\phi(\bm{r},t)$, surfactant concentration $c(\bm{r},t)$, polarization $\bm{p}(\bm{r},t)$, and fluid velocity $\bm{v}(\bm{r},t)$.
Unlike many previous models that introduce phenomenological terms to enforce stability or empirical adsorption laws, our formulation derives all equations from a single Rayleighian functional and a unified coarse-grained free energy, preserving detailed balance at equilibrium and ensuring thermodynamic consistency.

A key advantage of this framework is that Marangoni flows arise naturally without the need to impose an additional surface-tension–gradient term.
Because each surfactant exerts a tangential force on the interface, any spatial heterogeneity in surfactant concentration produces an excess stress that drives flow along the interface.
The resulting long-ranged disturbances in the fluid velocity are therefore the exact analogue of the classical Marangoni flows generated by gradients in surface tension.
Moreover, the model is thermodynamically consistent with both the Gibbs adsorption isotherm for surface-tension reduction and Henry’s law for the excess of surfactant at interfaces.

Assuming weak coupling between the fluid and surfactants allows us to obtain a perturbative equilibrium solution for a flat interface.
We validate this analytical solution by numerically simulating the same planar interface using a hybrid finite-difference–pseudospectral method.
This solution confirms that our model accurately captures the reduction in surface tension due to surfactant adsorption, in agreement with both previous phase-field models and experimental observations.
We then perform numerical simulations of an emulsion to demonstrate how the inclusion of the polarization field $\bm{p}(\bm{r},t)$ suppresses droplet coalescence.
Previous studies have used pseudopotential-based methods to generate ultra-stable emulsions~\cite{pelusi2022tlbfind}.
In contrast, our results show that stable emulsification can be achieved within a continuum framework, provided the surfactant polarization dynamics are explicitly incorporated.
  
%%%  
\section{Microscopic Derivation of the Hydrodynamic Equations \label{sec:hydrodynamics}}
%%%

At the microscopic level, each surfactant molecule is modelled as a dumbbell composed of two mass points: a hydrophilic head (H) and a hydrophobic tail (T), connected by a rigid rod of length $\ell$~\cite{kawakatsu_hybrid_1990}, see Fig~\ref{fig:surfactant-diagram}, right. 
The surrounding fluid is treated as a continuum, described by a velocity field $\bm{v}(\bm{r},t)$ and a volume fraction field $\phi(\bm{r},t)$. 
The field $\phi(\bm{r},t)$ represents the local volume fraction of one component (e.g., oil) relative to the total volume of both components (oil and water), with $\phi > 0$ indicating a local excess of component 1 and $\phi < 0$ indicating a local excess of component 2. 
In this work, we take component 1 to be oil and component 2 to be water, although this labelling is arbitrary. 
Under appropriate conditions, a diffuse oil–water interface emerges, to which the surfactant molecules are attracted.
In Fig.~\ref{fig:surfactant-diagram}, right, the vector $\hat{\bm{e}}_i$ denotes the orientation of surfactant molecule $i$, pointing from tail to head, with the centre of mass located at $\bm{r}_i$. 
The head is attracted to the aqueous phase, while the tail prefers the oil phase. 
These preferential interactions generate a net force $\bm{F}_i^{\text{fluid}}$ and a net torque $\bm{T}_i^{\text{fluid}}$ (about the centre of mass) when the molecule is not perpendicular with the interface or displaced from the midpoint of the interface.
From Fig.~\ref{fig:surfactant-diagram}, right, $\bm{F}_i^{\text{fluid}}$ and $\bm{T}_i^{\text{fluid}}$ can be expressed as:
\begin{subequations} \label{eq:force-torque}
    \begin{align}
    \bm{F}_i^{\text{fluid}} &= \bm{F}_\mathrm{H} \left( \bm{r}_i^+ \right) + \bm{F}_\mathrm{T}\left( \bm{r}_i^- \right), \\
    \bm{T}_i^{\text{fluid}} &= \frac{\ell}{2} \, \hat{\bm{e}}_i \times \bm{F}_\mathrm{H}\left( \bm{r}_i^+ \right) - \frac{\ell}{2}\, \hat{\bm{e}}_i \times \bm{F}_\mathrm{T}\left( \bm{r}_i^- \right),
\end{align}
\end{subequations}
where $\bm{F}_\mathrm{H} \left( \bm{r}_i^+ \right)$ and $\bm{F}_\mathrm{T} \left( \bm{r}_i^- \right)$ are the forces acting from the fluid on the surfactant mass points H and T, respectively, 
with $\bm{r}_i^\pm = \bm{r}_i \pm \ell\hat{\bm{e}}_i/2$ denotes the positions of H (for $+$) and T (for $-$).
The forces are obtained as:  
\begin{subequations} \label{eq:force-micro}
    \begin{align}
    \bm{F}_\mathrm{H} (\bm{r}_i^+) &= -\chi  \bm{\nabla}\phi(\bm{r})|_{\bm{r}=\bm{r}_i^+}  \, , \\
    \bm{F}_\mathrm{T} (\bm{r}_i^-) &= +\chi \bm{\nabla}\phi(\bm{r})|_{\bm{r}=\bm{r}_i^-} \, ,
\end{align}
\end{subequations}
where $\bm{\nabla}\phi(\bm{r})|_{\bm{r}=\bm{r}_i^\pm}$ is the spatial derivative of $\phi(\bm{r})$, evaluated at $\bm{r}_i^\pm$. 
In addition, $\chi>0$ is the interaction strength between each mass point and their preferential fluid phases (assumed to be equal in magnitude).
Substituting Eq.~\eqref{eq:force-micro} into \eqref{eq:force-torque}, and Taylor expanding for small $\ell$, we obtain the net force and the net torque acting on each molecule $i$ from the fluid:
\begin{subequations} \label{eq:force-torque-2}
    \begin{align}
    \bm{F}_i^{\text{fluid}} &= -\chi\ell(\hat{\bm{e}}_i\cdot\bm{\nabla}_i)\bm{\nabla}_i\phi(\bm{r}_i) + \mathcal{O}(\ell^2) \label{eq:total-force},\\
    \bm{T}_i^{\text{fluid}} &= -\chi\ell\hat{\bm{e}}_i\times\bm{\nabla}_i\phi(\bm{r}_i) + \mathcal{O}(\ell^2), \label{eq:total-torque}
\end{align}
\end{subequations}
where $\bm{\nabla}_i$ indicates derivative with respect to $\bm{r}_i$, centre of mass of molecule $i$.
Motivated by the structure of the force and torque expressions above, we derive the corresponding microscopic dynamics below.
From this microcopic dynamics, we can then obtain the coarse-grained hydrodynamic equations governing the coupled surfactant–binary fluid systems.

%%%
\subsection{Rayleighian dissipation for a single surfactant molecule in a binary fluid \label{sec:quiescent}}
%%%

In this section, we first derive the microscopic dynamics of a single surfactant molecule in a binary fluid using Rayleigh’s minimum energy dissipation principle~\cite{Doi_2011,doi2013soft}.
We then extend this framework to an ensemble of non-interacting surfactant molecules by introducing a single-particle distribution function.
This allows us to derive the coarse-grained hydrodynamic equations in terms of the concentration field $c(\bm{r},t)$, polarization field $\bm{p}(\bm{r},t)$, binary fluid volume fraction field $\phi(\bm{r},t)$ and velocity field $\bm{v}(\bm{r},t)$.

We first consider a system of a single dumbbell with two point masses at $\bm{r}^+$ and $\bm{r}^-$, representing the positions of the head and the tail respectively. 
We also have $|\bm{r}^+-\bm{r}^-|= \ell$, where $\ell$ is the length of the dumbbell.
The free energy of a single dumbbell particle in a continuum fluid is then given by:
\begin{equation}\label{eq:full_free_energy}
  A[\phi,\{\bm{r}^+,\bm{r}^-\}] = F_\text{fluid}[\phi] + \int d^dr \, \left\{ \chi\left[\delta(\bm{r}-\bm{r}^+) - \delta(\bm{r}-\bm{r}^-)\right] \phi(\bm{r}) \right\},
\end{equation}
where 
\begin{equation}
F_\text{fluid}[\phi]=\int d^dr \left\{ \frac{\alpha}{2}\phi^2 + \frac{\beta}{4}\phi^4 + \frac{\kappa}{2}|\nabla\phi|^2 \right\} 
\end{equation} 
is a typical free energy of a binary fluid, containing a double-well potential (for $\alpha<0$ and $\beta>0$) to drive bulk phase separation which competes with a squared-gradient term to penalise surface proliferation. 
Here we take $\alpha=-\beta<0$, so that equilibrium phases correspond to $\phi=\pm\sqrt{-\alpha/\beta}=\pm1$.
Note that $d$ is the spatial dimension, and the integral is taken over the volume of the system.
The second term in Eq.~(\ref{eq:full_free_energy}) couples the two discrete head and tail positions to the volume fraction field, with each particle-field interaction characterised by an interaction strength $\chi$~\cite{hardy2024hybrid}. 
We now introduce the orientation unit vector $\hat{\bm{e}}$ and the centre of mass position $\bm{R}$ to be:
\begin{align}
    \hat{\bm{e}} = \frac{\bm{r}^+-\bm{r}^-}{\ell} \quad\text{and}\quad  \bm{R} &= \frac{\bm{r}^++\bm{r}^-}{2}.
\end{align}
Then, Eq.~(\ref{eq:full_free_energy}) becomes:
\begin{equation} \label{eq:free_energy_1}
  A[\phi,\{\bm{R},\hat{\bm{e}}\}] = F_\text{fluid}[\phi]
  + \int d^dr \left\{ \chi\left[\delta\left(\bm{r}-\left(\bm{R}+\frac{\ell}{2}\hat{\bm{e}}\right)\right) 
  				    - \delta\left(\bm{r}-\left(\bm{R}-\frac{\ell}{2}\hat{\bm{e}}\right)\right) \right] \phi(\bm{r}) \right\}.
\end{equation}
Taylor expanding Eq.~(\ref{eq:free_energy_1}) for small $\ell$, we may obtain:
\begin{equation} \label{eq:free_energy_2}
  A[\phi,\{\bm{R},\hat{\bm{e}}\}] = F_\text{fluid}[\phi] + \chi\ell(\hat{\bm{e}}\cdot\bm{\nabla}_{\bm{R}})\phi(\bm{R}) + \mathcal{O}(\ell^3),
\end{equation}
where $\bm{\nabla}_{\bm{R}}$ denotes spatial derivative with respect to the centre of mass of the dumbbell $\bm{R}$.
Thus the hybrid discrete-particle-continuum-fluid free energy is a functional of the volume fraction field $\phi(\bm{r})$ and a function of discrete variables $\bm{R}$ and $\hat{\bm{e}}$.
Ultimately, we aim to re-cast Eq.~(\ref{eq:free_energy_1}) or (\ref{eq:free_energy_2}) as a purely continuum free energy.

Following~\cite{Doi_2011}, we next write down the Rayleigh dissipation functional
\begin{equation}\label{eq:rayleighian_full}
  \mathcal{R}[\partial_t\phi,\bm{v}_1,\bm{v}_2,\{\dot{\bm{r}}^+,\dot{\bm{r}}^-\}] = \Phi_1[\bm{v}] + \Phi_2[\bm{v}_1,\bm{v}_2] + \Phi_3[\bm{v},\{\dot{\bm{r}}^+,\dot{\bm{r}}^-\}]+ \dot{A}
\end{equation}
which is a functional of the rate of change of the volume fraction $\partial_t\phi$, the velocity of fluid component 1: $\bm{v}_1(\bm{r},t)$, the velocity of fluid component 2: $\bm{v}_2(\bm{r},t)$, and the velocities of the discrete particles $\{\dot{\bm{r}}^+,\dot{\bm{r}}^-\}$. 
We may also further define the total velocity of the fluid to be 
\begin{equation}
\bm{v}(\bm{r},t)=\frac{1+\phi}{2}\bm{v}_1(\bm{r},t)+\frac{1-\phi}{2}\bm{v}_2(\bm{r},t),
\end{equation} 
so that either $\bm{v}_1$ or $\bm{v}_2$ may be expressed in terms of $\bm{v}$ and the other.
We use a dot over any variable to denote its total time-derivative. 
We neglect the volume of discrete particles in this formulation. 
The first dissipation term in Eq.~(\ref{eq:rayleighian_full}) denotes the viscous dissipation of the incompressible Newtonian fluid,
\begin{equation}
  \Phi_1[\bm{v}] = \int d^dr \, \frac{\eta}{4} \left[\bm{\nabla}\bm{v}+(\bm{\nabla}\bm{v})^T\right]^2 - \int d^dr \, P(\bm{r}) \bm{\nabla}\cdot\bm{v},
\end{equation}
where $\eta$ is the fluid viscosity (assumed to be the same for both components of the fluid) and $P(\bm{r})$ is a Lagrange multiplier to enforce incompressibility condition $\bm{\nabla}\cdot\bm{v}=0$, which also happens to be the pressure of the fluid.
The second dissipation term in Eq.~(\ref{eq:rayleighian_full}) represents dissipation due to velocity differences between the two fluid components,
\begin{equation}
  \Phi_2[\bm{v}_1,\bm{v}_2] =\int d^dr \, \frac{\xi}{2} (\bm{v}_1-\bm{v}_2)^2 =  \int d^dr \, \frac{2\xi}{(1-\phi)^2}(\bm{v}_1-\bm{v})^2,
\end{equation}
where $\xi>0$ is a phenomenological dissipation parameter related to the mobility. 
The third term in Eq.~(\ref{eq:rayleighian_full}) couples the fluid velocity to the particle velocities
\begin{equation} \label{eq:Phi-3}
    \Phi_3[\bm{v},\{\dot{\bm{r}}^+,\dot{\bm{r}}^-\}] = -\int d^dr 
    \left\{  \delta(\bm{r}-\bm{r}^+)\bm{f}^+\cdot(\bm{v}(\bm{r})-\dot{\bm{r}}^+) 
          +  \delta(\bm{r}-\bm{r}^-)\bm{f}^-\cdot(\bm{v}(\bm{r})-\dot{\bm{r}}^-) \right\}
\end{equation}
where $\bm{f}^\pm$ are constraint forces which enforce $\bm{v}(\bm{r}^\pm)=\dot{\bm{r}}^\pm$ (similar to Lagrange multipliers). 
This is equivalent to introducing point forces acting on the fluid (i.e., Stokeslet solution).
We can define the angular velocity of the dumbbell $\bm{\omega}$ to be $\dot{\hat{\bm{e}}}=\bm{\omega}\times\hat{\bm{e}}$.
The rate of change of the positions $\dot{\bm{r}}^\pm$ can then be written in terms of $\dot{\bm{R}}$ and $\bm{\omega}$:
\begin{equation}\label{eq:rdot}
\dot{\bm{r}}^\pm = \dot{\bm{R}} \pm \bm{\omega}\times\frac{\ell}{2}\hat{\bm{e}}.
\end{equation}
Using Eq.~(\ref{eq:rdot}), Eq.~(\ref{eq:Phi-3}) can then be written in terms of $\dot{\bm{R}}$ and $\bm{\omega}$, and thus $\Phi_3[\bm{v},\{\dot{\bm{R}},\bm{\omega}\}]$ is now a function of $\dot{\bm{R}}$ and $\bm{\omega}$.
The final term in Eq.~(\ref{eq:rayleighian_full}) is simply the derivative of the free energy with time,
\begin{equation}\label{eq:dot_free_energy}
  \begin{split}
    \dot{A}[\partial_t\phi,\{\dot{\bm{R}},\bm{\omega}\}] = & \int d^dr \left\{ \frac{\delta F_\text{fluid}}{\delta \phi} \partial_t\phi \right\}
    + \int d^dr \left\{ \chi\partial_t\phi
    \left[ \delta\left(\bm{r}-\left(\bm{R}+\frac{\ell}{2}\hat{\bm{e}}\right)\right) - \delta\left(\bm{r}-\left(\bm{R}-\frac{\ell}{2}\hat{\bm{e}}\right)\right) \right] \right\} \\
    & +\chi \left[ \bm{\nabla}\phi(\bm{r})|_{\bm{r}=\bm{R}+\ell\hat{\bm{e}}/2} - \bm{\nabla}\phi(\bm{r})|_{\bm{r}=\bm{R}-\ell\hat{\bm{e}}/2} \right] \cdot\dot{\bm{R}}\\
    & +\frac{\ell\chi}{2} \left[ \bm{\nabla}\phi(\bm{r})|_{\bm{r}=\bm{R}+\ell\hat{\bm{e}}/2} + \bm{\nabla}\phi(\bm{r})|_{\bm{r}=\bm{R}-\ell\hat{\bm{e}}/2} \right] \cdot \bm{\omega}\times\hat{\bm{e}},
  \end{split}
\end{equation}
where we have expressed $\dot{\hat{\bm{e}}}=\bm{\omega}\times\hat{\bm{e}}$.
Minimizing the Rayleighian  $\mathcal{R}[\partial_t\phi,\bm{v},\bm{v}_1,\{\dot{\bm{R}},\bm{\omega}\}]$ with respect to the generalized velocities $\partial_t\phi$, $\bm{v}$, $\bm{v}_1$, $\dot{\bm{R}}$ and $\bm{\omega}$, and subject to the constraint that volume fraction is conserved, i.e. 
\begin{equation}\label{eq:continuity_phi}
    \partial_t\phi + \bm{\nabla}\cdot [(1+\phi)\bm{v}_1] = 0,
\end{equation}
we get the equations of motion for the system:
\begin{subequations} \label{eq:hybrid_dynamics}
  \begin{align}
    \partial_t\phi  = &-\bm{v}\cdot\bm{\nabla}\phi + M\nabla^2\bigg(\frac{\delta A}{\delta \phi}\bigg) \label{eq:phi_dynamics_dipole}, \\
    \bm{v}(\bm{r}) = & \int d^dr' \left[ \bm{O}(\bm{r}-\bm{r}') \cdot 
    			\left(-\phi(\bm{r}')\bm{\nabla}'\frac{\delta A}{\delta \phi(\bm{r}')}\right) \right]
    + \left[ \bm{O}(\bm{r}-\bm{R}) + \frac{\ell^2}{8}(\hat{\bm{e}}\hat{\bm{e}}:\bm{\nabla}_{\bm{R}}\bm{\nabla}_{\bm{R}})
    				\bm{O}(\bm{r}-\bm{R})\right] \cdot\bm{g}  \nonumber\\
    &+ \left[ (\bm{e}\cdot\bm{\nabla}_{\bm{R}})\bm{O}(\bm{r}-\bm{R}) + 
    			\frac{\ell^2}{3!}(\hat{\bm{e}}\hat{\bm{e}}\hat{\bm{e}}\fcolon\bm{\nabla}_{\bm{R}}\bm{\nabla}_{\bm{R}}\bm{\nabla}_{\bm{R}})\bm{O}(\bm{r}-\bm{R}) \right]\cdot\bm{h}+\mathcal{O}(\ell^3)\label{eq:velocity_dipole}\\
    \bm{g}&=-\ell\chi(\hat{\bm{e}}\cdot\bm{\nabla}_{\bm{R}})\bm{\nabla}_{\bm{R}}\phi(\bm{R}) + \mathcal{O}(\ell^3),\\
    \bm{h}&=-\ell\chi\bm{\nabla}_{\bm{R}}\phi(\bm{R}) + \mathcal{O}(\ell^3),\\
    \dot{\bm{R}}&= \bm{v}(\bm{R})+\mathcal{O}(\ell^2),\\
    \dot{\hat{\bm{e}}}
    &= (\hat{\bm{e}}\cdot\bm{\nabla}_{\bm{R}})\bm{v}(\bm{R})+\mathcal{O}(\ell^2),
  \end{align}
\end{subequations}
where we have defined $\bm{g} = \bm{f}_{2}+\bm{f}_{1}$ and $\bm{h} = \ell(\bm{f}_{2}-\bm{f}_{1})/2$.
$\bm{O}(\bm{r})$ is the second rank Oseen tensor, which in three-dimension ($d=3$), is given by:
\begin{equation}
    \bm{O}(\bm{r})=\frac{1}{8\pi\eta r}\bigg(\bm{I}+\frac{\bm{r}\bm{r}}{r^2}\bigg),
\end{equation}
where $\bm{I}$ is the identity matrix.
To obtain Eqs.~(\ref{eq:hybrid_dynamics}), we have assumed that the binary fluid dissipation coefficient, $\xi \equiv (1+\phi)^2(1-\phi)^2 / (4M)$, depends explicitly on the volume fraction, so that the mobility $M$ in Eq.~(\ref{eq:phi_dynamics_dipole}) remains constant.

The singular nature of $\bm{O}(\bm{r})$ at $\bm{r}=\bm{0}$ poses a problem, 
as the centre-of-mass and orientation dynamics both require evaluating $\bm{v}(\bm{R})$. 
However, there is a precedent in the literature to remedy this, particularly in polymer physics when one deals with strings of spherical beads \cite{doi1986theory}. 
One can begin by assuming the singular term in the velocity dynamics is omitted. 
We will denote the remaining external flow velocity as $\bm{v}_{\text{ext}}(\bm{r})$, defined simply as
\begin{equation}
  \begin{split}\label{eq:velocity_external_flow}
    \bm{v}_\text{ext}(\bm{r}) = &\int d^dr' \left[ \bm{O}(\bm{r}-\bm{r}')\cdot
    		\left(-\phi(\bm{r}')\bm{\nabla'}\frac{\delta A}{\delta \phi(\bm{r}')} \right) \right]
    +\mathcal{O}(\ell^3).
  \end{split}
\end{equation}
To compensate for the neglect of the singular self-interaction flow, one needs to supplement the equations of $\dot{\bm{R}}$ and $\dot{\hat{\bm{e}}}$ by a self-interaction term, which is taken ad-hoc to be the mobility matrix of the isolated particle multiplied by the force on that particle (neglecting hydrodynamic forces). 
For us, we choose the mobility matrix which is consistent with an isolated needle-like particle, as that is the closest to our dipolar case for which an exact solution is available. Then,
\begin{subequations} \label{eq:dipole_dynamics}
  \begin{align}
  \dot{\hat{\bm{e}}} = & \left(\bm{I}- \hat{\bm{e}}\hat{\bm{e}}\right) \cdot
   \left[ (\hat{\bm{e}}\cdot\bm{\nabla}_{\bm{R}})\bm{v}_{\text{ext}}(\bm{R}) - \frac{1}{\gamma_r}\frac{\partial A}{\partial \hat{\bm{e}}} \right],
  \label{eq:orientation_dipole}\\
    \dot{\bm{R}}
    = &\bm{v}_{\text{ext}}(\bm{R}) - \left[ \frac{1}{\zeta_{\parallel}}  \hat{\bm{e}}\hat{\bm{e}}
    + \frac{1}{\zeta_{\perp}} \left( \bm{I} - \hat{\bm{e}}\hat{\bm{e}} \right)  \right]
    \cdot\frac{\partial A}{\partial \bm{R}}. \label{eq:position_dipole}
  \end{align}
\end{subequations}
The operator $(\bm{I}-\hat{\bm{e}}\hat{\bm{e}})$ is known as the perpendicular projection operator, as it projects any vector $\bm{V}$ onto the plane perpendicular to $\hat{\bm{e}}$, i.e., $(\bm{I}-\hat{\bm{e}}\hat{\bm{e}})\cdot\bm{V}$. 
We employ this operator in Eq.~\eqref{eq:orientation_dipole} to ensure that the unit-length constraint $|\hat{\bm{e}}|=1$ is maintained throughout the dynamics.
In Eqs.~\eqref{eq:dipole_dynamics},  $\gamma_r$ is the rotational friction, $\zeta_{\parallel}$ and $\zeta_{\perp}$ are translational friction parallel and perpendicular to $\hat{\bm{e}}$ respectively. 
In the case of ellipsoidal particles, these friction coefficients can be written in terms of microscopic parameters~\cite{Hoffmann2009,kim2005microhydrodynamics}. 
This result reproduces the Jeffrey orbits if we define the symmetric and anti-symmetric tensors
\begin{subequations}\label{eqs:velocity_tensors}
  \begin{align}
    (\bm{D}_{\text{ext}})_{\alpha\beta} \equiv& \frac{1}{2}
    	\left[(\bm{\nabla})_{\alpha}(\bm{v}_{\text{ext}})_{\beta} + (\bm{\nabla})_{\beta}(\bm{v}_{\text{ext}})_{\alpha}\right],\\
    (\bm{\Omega}_{\text{ext}})_{\alpha\beta} \equiv & \frac{1}{2}
    	\left[(\bm{\nabla})_{\alpha}(\bm{v}_{\text{ext}})_{\beta}-(\bm{\nabla})_{\beta}(\bm{v}_{\text{ext}})_{\alpha}\right],
  \end{align}
\end{subequations}
i.e.
\begin{equation}\label{eq:orientation_dipole_jeffrey}
  \dot{\hat{\bm{e}}}
  = \underbrace{-\bm{\Omega}_{\text{ext}}(\bm{R}) \cdot\hat{\bm{e}}
  +B\left[\bm{D}_{\text{ext}}(\bm{R}) \cdot \hat{\bm{e}} - \left(\hat{\bm{e}}\cdot\bm{D}_{\text{ext}}(\bm{R}) \cdot \hat{\bm{e}}\right)\hat{\bm{e}}\right]}_{ \left(\bm{I}- \hat{\bm{e}}\hat{\bm{e}}\right) \cdot (\hat{\bm{e}}\cdot\bm{\nabla}_{\bm{R}}) \bm{v}_\text{ext}(\bm{R}) \text{ for } B=1}
  - \left(\bm{I}-\hat{\bm{e}}\hat{\bm{e}}\right) \cdot \bigg(\frac{1}{\gamma_r}\frac{\partial A}{\partial \hat{\bm{e}}}\bigg),
\end{equation}
though we specifically have $B=1$ since we assumed the surfactant had a needle-like aspect ratio in our derivation. 
In the case of ellipsoidal particles, $B$ is related to the aspect ratio of the particle $\Delta$ via $B=(\Delta^2-1)/(\Delta^2+1)$~\cite{jeffery1922motion}.
Up to this point we have derived the dynamics of a single surfactant molecule in the presence of a phase-separating binary fluid, Eqs.~(\ref{eq:orientation_dipole}), (\ref{eq:position_dipole}) and (\ref{eq:orientation_dipole_jeffrey}), 
subject to external flow due to phase separation, Eq.~(\ref{eq:velocity_external_flow}).

Finally, we introduce noise to the single particle dynamics to get the Ito stochastic differential equations (in $d$-dimension),
\begin{subequations} \label{eq:ito-dynamics}
  \begin{align} 
    d\bm{R} = &\left\{ \bm{v}_{\text{ext}}(\bm{R})
    	-\left[ \frac{1}{\zeta_{\parallel}} \hat{\bm{e}}\hat{\bm{e}} + \frac{1}{\zeta_{\perp}}\left( \bm{I}- \hat{\bm{e}}\hat{\bm{e}} \right)\right]
    \cdot\frac{\partial A}{\partial \bm{R}} \right\}dt 
    +  \left[ \sqrt{\frac{2k_\mathrm{B}T}{\zeta_{\parallel}}} \hat{\bm{e}}\hat{\bm{e}} + \sqrt{\frac{2k_\mathrm{B}T}{\zeta_{\perp}}}
    \left( \bm{I}-\hat{\bm{e}}\hat{\bm{e}} \right) \right] \cdot d\bm{W}_{\bm{R}}, \label{eq:ito_position}\\
    d\hat{\bm{e}} = &\left(\bm{I}-\hat{\bm{e}}\hat{\bm{e}}\right)\cdot
    		\left\{  \left[ (\hat{\bm{e}}\cdot\bm{\nabla}_{\bm{R}})\bm{v}_{\text{ext}}(\bm{R}) 
				- \frac{1}{\gamma_r}\frac{\partial A}{\partial\hat{\bm{e}}} \right] dt
    +\sqrt{\frac{2k_\mathrm{B}T}{\gamma_r}}d\bm{W}_{\bm{e}} \right\}
    -\frac{(d-1)k_\mathrm{B}T}{\gamma_r}\hat{\bm{e}} \, dt,
    \label{eq:ito_orientation}
  \end{align}
\end{subequations}
where $d\bm{W}_{\bm{R}}$ and $d\bm{W}_{\bm{e}}$ are Gaussian white noise with zero mean and variance $dt$.
If one instead wanted the Stratonovich form of the equations, the last term (i.e. the drift) is omitted in Eq.~(\ref{eq:ito_orientation}) above.
Using the form of the free energy $A[\phi,\{\bm{R},\hat{\bm{e}}\}]$ in Eq.~\eqref{eq:free_energy_2}, we can calculate (up to order $\ell^2$):
\begin{subequations} \label{eq:A-deriv}
  \begin{align} 
  \frac{\partial A}{\partial \bm{R}} &= \chi\ell(\hat{\bm{e}}\cdot\bm{\nabla}_{\bm{R}})\bm{\nabla}_{\bm{R}}\phi(\bm{R}) = -\bm{F}^{\text{fluid}}, \\
  \frac{\partial A}{\partial \hat{\bm{e}}} &= \chi\ell\bm{\nabla}_{\bm{R}}\phi(\bm{R}) = \underbrace{\chi\ell\left[\hat{\bm{e}}\times\bm{\nabla}_{\bm{R}}\phi(\bm{R})\right]}_{-\bm{T}^{\text{fluid}}}\times\hat{\bm{e}} 
  + \chi\ell\left[\hat{\bm{e}}\cdot\bm{\nabla}_{\bm{R}}\phi(\bm{R})\right]\hat{\bm{e}}. \label{eq:dAde}
  \end{align}
\end{subequations}
Note that the last term in Eq.~\eqref{eq:dAde} does not affect the dynamics due to the perpendicular projection operator $\bm{I}-\hat{\bm{e}}\hat{\bm{e}}$ in Eq.~\eqref{eq:ito_orientation}.
Thus, we identify the derivatives $\partial A/\partial\bm{R}$ and $\partial A/\partial\hat{\bm{e}}$ as being directly related to the net force $\bm{F}^{\text{fluid}}$ and net torque $\bm{T}^{\text{fluid}}$ exerted by the fluid on the molecule [cf. Eqs.~\eqref{eq:force-torque-2}]. 
Accordingly, the deterministic parts of Eqs.~\eqref{eq:ito-dynamics} in the absence of external flow become:
\begin{subequations} 
  \begin{align} 
  \frac{d\bm{R}}{dt} &= \left[ \frac{1}{\zeta_{\parallel}} \hat{\bm{e}}\hat{\bm{e}} + \frac{1}{\zeta_{\perp}}\left( \bm{I}- \hat{\bm{e}}\hat{\bm{e}} \right)\right]\cdot\bm{F}^{\text{fluid}}, \\
  \frac{d\hat{\bm{e}}}{dt} &= \frac{1}{\gamma_r}\bm{T}^{\text{fluid}}\times\hat{\bm{e}},
  \end{align}
\end{subequations}
as expected from overdamped dynamics.

%%%
\subsection{Coarse-graining the single surfactant particle dynamics\label{sec:coarse-graining}}
%%%

The dynamics described in Eqs.~(\ref{eq:ito_position}) and (\ref{eq:ito_orientation}) in terms of surfactant particle position $\bm{R}$ and orientation $\hat{\bm{e}}$ can be equivalently described by the Smoluchowski equation \cite{Doi_2011,doi1986theory}:
\begin{equation}\label{eq:smol}
  \frac{\partial \psi}{\partial t}
  = -\bm{\nabla}\cdot\bm{J}_{\bm{r}}
  +\bm{\mathcal{R}}\cdot\bigg(\frac{k_\mathrm{B}T}{\gamma_r}\bm{\mathcal{R}}\psi
  +\frac{1}{\gamma_r}\psi\bm{\mathcal{R}}A
  -\hat{\bm{e}}\times\bm{K}\cdot\hat{\bm{e}}\psi\bigg),
\end{equation}
where $\psi(\bm{r},\hat{\bm{e}},t)$ is the probability density for the surfactant particle.
More precisely, $\psi(\bm{r},\hat{\bm{e}},t)d^drd^{d-1}\hat{e}$ is the probability of finding a surfactant particle inside an infinitesimal volume $d^dr$, located at $\bm{r}$, with the orientation pointing in the direction of the solid angle $d^{d-1}\hat{e}$, located at $\hat{\bm{e}}$.
In Eq.~(\ref{eq:smol}), $\bm{K}$ is defined to be $\bm{K}\equiv(\bm{\nabla}\bm{v}_\text{ext})^T$, or
$(\bm{K})_{\alpha\beta}\equiv (\bm{\nabla})_{\beta}(\bm{v}_{\text{ext}})_{\alpha}$.
We have also define the angular derivative operator $\bm{\mathcal{R}}$ to be:
\begin{equation}
    (\bm{\mathcal{R}})_{\alpha} \equiv \left( \hat{\bm{e}}\times\frac{\partial}{\partial \hat{\bm{e}}} \right)_{\alpha}
    = \varepsilon_{\alpha\beta\gamma}(\hat{\bm{e}})_{\beta} \left( \frac{\partial}{\partial\hat{\bm{e}}} \right)_{\gamma}.
\end{equation}
$\bm{J}_{\bm{r}}$ in Eq.~(\ref{eq:smol}) is the positional flux, 
\begin{equation}
    \begin{split}
    \bm{J}_{\bm{r}} \equiv& \bigg\{ \bm{v}_{\text{ext}}(\bm{r}) - 
    \left[  \frac{1}{\zeta_{\parallel}} \hat{\bm{e}} \hat{\bm{e}} + \frac{1}{\zeta_{\perp}}
    			\left(\bm{I}  -   \hat{\bm{e}} \hat{\bm{e}} \right)  \right]
    	\cdot\bm{\nabla} A\bigg\} \psi
    - k_\mathrm{B}T \left[  \frac{1}{\zeta_{\parallel}}  \hat{\bm{e}} \hat{\bm{e}}
    		+\frac{1}{\zeta_{\perp}} \left(\bm{I} -  \hat{\bm{e}} \hat{\bm{e}} \right)  \right]
    	\cdot\bm{\nabla}\psi.\label{eq:smol_position_flux} \end{split}
\end{equation}
In the absence of an external velocity $\bm{v}_{\text{ext}}(\bm{r})=\bm{0}$, one can show the steady state solution to the Smoluchowski equation (\ref{eq:smol}) and the $\phi$-dynamics (\ref{eq:phi_dynamics_dipole}) is the Boltzmann distribution:
\begin{equation}\label{eq:psi_boltzmann_single}
  \psi(\bm{r},\hat{\bm{e}},t\to\infty) \propto e^{-\frac{ A[\phi(\bm{r}),\hat{\bm{e}}]}{k_\mathrm{B}T}} \quad\text{and}\quad 
  \frac{\delta A}{\delta\phi} = 0,
\end{equation}
where $A$ is the single-molecule free energy from Eq.~(\ref{eq:free_energy_2}):
\begin{equation} \label{eq:free_energy_3}
    A[\phi(\bm{r}),\hat{\bm{e}}] = F_\text{fluid}[\phi] + \chi\ell\hat{\bm{e}}\cdot\bm{\nabla}\phi(\bm{r}). \\
\end{equation}
Thus, our hybrid discrete-particle-continuum-fluid model satisfies the detailed balance condition in the steady state, i.e. equilibrium.

To derive the local concentration of surfactant particles $c(\bm{r},t)$ and the local average orientation of the surfactant particles $\bm{p}(\bm{r},t)$, we expand the distribution function $\psi$ as in Appendix \ref{app:distribution_expansion}. 
Inserting this into Eq.~(\ref{eq:smol}) and using the expression of the free energy in Eq.~(\ref{eq:free_energy_3}),
we can write
\begin{equation}
  \begin{split}
    \frac{\partial c(\bm{r},t)}{\partial t} =& -\bm{\nabla} \cdot \bigg\{
    	\bm{v}_{\text{ext}}(\bm{r})c(\bm{r},t) - \frac{\chi\ell c(\bm{r},t)}{d+2} 
		\left[ \left(\frac{1}{\zeta_{\parallel}}-\frac{1}{\zeta_{\perp}} \right) \bm{p}(\bm{r},t)\nabla^2\phi(\bm{r},t)
    			+ \left( \frac{2}{\zeta_{\parallel}}+\frac{d}{\zeta_{\perp}} \right) (\bm{p}(\bm{r},t)\cdot\bm{\nabla})\bm{\nabla}\phi(\bm{r},t) \right]
    \\
    & -\frac{k_\mathrm{B}T}{d}\text{Tr}\big(\bm{\zeta}^{-1}\big) \bm{\nabla} c(\bm{r},t) \bigg\}
  \end{split}
\end{equation}
and
\begin{equation}
  \begin{split}
    \frac{\partial (c(\bm{r},t) \bm{p}(\bm{r},t))}{\partial t} =& -\bm{\nabla}\cdot \bigg\{
    	\bm{v}_{\text{ext}} (c \bm{p})  - \chi\ell \left( \frac{1}{\zeta_{\parallel}}-\frac{1}{\zeta_{\perp}} \right)
    	\frac{c}{d(d+2)}  \left( \bm{I}\nabla^2\phi + 2\bm{\nabla}\bm{\nabla}\phi \right)
    	-\frac{\chi\ell}{\zeta_{\perp}} \frac{c(\bm{r},t)}{d}\bm{I}\nabla^2\phi \\
    & -\frac{k_\mathrm{B}T}{d+2} \left( \frac{1}{\zeta_{\parallel}}-\frac{1}{\zeta_{\perp}} \right)
    	\left[ \bm{\nabla}(c\bm{p}) + \big(\bm{\nabla}(c\bm{p})\big)^T + \bm{I}\bm{\nabla}\cdot(c\bm{p}) \right] 
      -\frac{k_\mathrm{B}T}{\zeta_{\perp}}\big(\bm{\nabla}(c\bm{p})\big)^T \bigg\} \\
    &-\frac{k_\mathrm{B}T(d-1)}{\gamma_r}c\bm{p}
    -\frac{\chi\ell}{\gamma_r}\frac{d-1}{d}c\bm{\nabla}\phi 
    +B\frac{d}{d+2}c\,\bm{D}_{\text{ext}}\cdot\bm{p} - c\,\bm{\Omega}_{\text{ext}}\cdot\bm{p}
  \end{split}
\end{equation}

If one considers a hydrodynamic length scale $\xi$, then it is possible to determine the relative contributions of rotational and translational diffusion on this length scale. For rod-like particles,
\begin{equation}
  \frac{\xi^2 \zeta_{\parallel}}{\gamma_r} \sim \bigg(\frac{\xi}{\ell}\bigg)^2.
\end{equation}
Therefore, when considering a length scale $\xi\gg\ell$, rotational diffusion is much faster than translational diffusion. 
On this length scale and in dimension $d$, the polarization dynamics simplifies to:
\begin{equation} \label{eq:polar_dynamics_external}
  \begin{split}
    \frac{\partial \bm{p}(\bm{r},t)}{\partial t} + \bm{v}_{\text{ext}}(\bm{r})\cdot\bm{\nabla}\bm{p}(\bm{r},t)
    = - \frac{d-1}{d\gamma_r c(\bm{r},t)}\frac{\delta F}{\delta\bm{p}}
     + \left[ B\frac{d}{d+2}\bm{D}_{\text{ext}}(\bm{r})-\bm{\Omega}_{\text{ext}}(\bm{r}) \right] \cdot \bm{p}(\bm{r},t)
  \end{split}
\end{equation}
and the concentration dynamics (taking $\zeta_{\perp}\approx\zeta_{\parallel}=\gamma_t$) are
\begin{equation} \label{eq:concentration_dynamics_external}
  \begin{split}
    \frac{\partial c(\bm{r},t)}{\partial t}
    = & -\bm{\nabla}\cdot\left[ \bm{v}_{\text{ext}}(\bm{r})c(\bm{r},t)
    	- \frac{1}{\gamma_t} c(\bm{r},t)\bm{\nabla}\left(\frac{\delta F}{\delta c}\right) \right]
  \end{split}
\end{equation}
with the identification of a coarse-grained mesoscopic Helmholtz free energy:
\begin{equation} \label{eq:coarse_grained_fe}
  F[\phi,c,\bm{p}] = \int d^dr \left[ \frac{\alpha}{2}\phi^2 + \frac{\beta}{4}\phi^4 + \frac{\kappa}{2}|\bm{\nabla}\phi|^2
  			  + k_\mathrm{B}T c\ln(a^dc) + \chi\ell c\bm{p}\cdot\bm{\nabla}\phi +\frac{d}{2}k_\mathrm{B}T c|\bm{p}|^2 \right]
\end{equation}
where $a$ is the typical microscopic lengthscale, introduced to make the term inside the logarithm dimensionless (the addition term $c\ln(a^d)$ drops out in the dynamics). 
The first two terms, $\alpha\phi^2/2$ and $\beta\phi^4/4$, are standard Cahn-Hilliard terms derived from thermodynamic theory~\cite{lee2014physical}, where $\alpha$ and $\beta$ are constants that characterize the behavior of the two phases. 
These terms form a double well potential representing the water and oil phases. 
Setting $\alpha<0$ and $\beta>0$ ensures that the two fluids remain distinct and do not mix. 
The third term, $\kappa|\bm{\nabla}\phi|^2/2$, is a semi-local term responsible for creating the diffuse water-oil interface, with $\kappa$ controlling the width of the interface. 
In particular, the width of the diffuse interface is given by $\xi_{\text{I}}=\sqrt{-2\kappa/\alpha}$~\cite{cates2018theories}.
The logarithmic term $k_\mathrm{B}T c \ln(a^d c)$ captures the translational diffusion of the surfactant molecules. 
Similarly, the term $dk_\mathrm{B}Tc |\bm{p}|^2/2$ accounts for the rotational diffusion. 
The term $\chi \ell c \bm{p} \cdot \bm{\nabla} \phi$ couples the surfactants to the fluid, causing them to be attracted and align perpendicular to the interface. 
In the absence of surfactant, $c=0$ or $\chi=0$, the bare surface tension of the water-oil interface is given by $\sigma_{\text{I}}=\sqrt{-8\kappa\alpha^3/(9\beta^2)}$.
Eq.~(\ref{eq:coarse_grained_fe}) can alternatively be derived from direct calculation of the Shannon entropy and energy contributions to the Helmholtz free energy, as shown in Appendix \ref{app:alt_deriv}.

We emphasize that the coarse-grained mesoscopic free energy $F$ is distinct from the hybrid single-molecule–continuum-fluid free energy $A$ in Eq.(\ref{eq:free_energy_2}), which depends explicitly on the discrete particle position. 
In the absence of external flow, $\bm{v}_{\text{ext}} = 0$, one can show that the steady-state solution to the $\bm{p}$- and $c$-dynamics in Eqs.(\ref{eq:polar_dynamics_external}) and~(\ref{eq:concentration_dynamics_external}), respectively, is given by the minimum of the coarse-grained mesoscopic free energy:
\begin{equation}
\frac{\delta F}{\delta c} = 0 \quad\text{and}\quad \frac{\delta F}{\delta\bm{p}} = \bm{0}.
\end{equation}
This shows that the detailed balance property is preserved under coarse-graining: 
at the single-particle level, the equilibrium steady state is characterized by a Boltzmann distribution [see Eq.~(\ref{eq:psi_boltzmann_single})], 
while at the coarse-grained level, it corresponds to the minimum of the mesoscopic free energy $F$.

%%%
\subsection{Coupling surfactant particle dynamics back to the fluid flow.}
%%%

The results of the previous subsection neglect hydrodynamic effects due to the presence of the surfactants, as can be seen in the definition of $\bm{v}_{\text{ext}}(\bm{r})$ in Eq.~(\ref{eq:velocity_external_flow}). 
To derive hydrodynamic flows from the surfactants, we again use the Rayleigh dissipation formalism. 
However, we now start with the coarse-grained free energy of Eq.~(\ref{eq:coarse_grained_fe}) in our definition of the Rayleighian, so that [c.f. Eq.~(\ref{eq:rayleighian_full})]
\begin{equation}\label{eq:rayleighian_cg}
    \mathcal{R}_{\text{cg}}[\partial_t\phi,\bm{v},\bm{v}_1,\partial_tc,\partial_t\bm{p}] = \Phi_1[\bm{v}] + \Phi_2[\bm{v}_1,\bm{v}_2] + \dot{F}
\end{equation}
where
\begin{equation}
    \dot{F} = \int d^3 r\bigg(\frac{\delta F}{\delta \phi}\partial_t\phi
    +\frac{\delta F}{\delta c}\partial_tc
    +\frac{\delta F}{\delta \bm{p}}\partial_t\bm{p}\bigg).
\end{equation}
$\partial_t\phi$, $\partial_tc$, and $\partial_t\bm{p}$ are defined in Eqs.~(\ref{eq:phi_dynamics_dipole}), (\ref{eq:concentration_dynamics_external}), and (\ref{eq:polar_dynamics_external}). 
Within these equations, we replace the external field $\bm{v}_{\text{ext}}\to\bm{v}$, where $\bm{v}(\bm{r},t)$ is now the total fluid velocity to be determined self-consistently, so
\begin{subequations} \label{eq:dynamics}
\begin{align}
    \frac{\partial\phi(\bm{r},t)}{\partial t} + \bm{\nabla}\cdot \left[\phi(\bm{r},t)\bm{v}(\bm{r},t) \right] 
    		&= M\nabla^2\frac{\delta F}{\delta \phi}, \label{eq:phi_continuity}\\
    \frac{\partial c(\bm{r},t)}{\partial t}  + \bm{\nabla}\cdot\left[ c(\bm{r},t)\bm{v}(\bm{r},t) \right]  
    		&=  \bm{\nabla}\cdot \left[ \frac{1}{\gamma_t}c(\bm{r},t)\bm{\nabla}\left(\frac{\delta F}{\delta c}\right) \right],\label{eq:concentration_dynamics}\\
    \frac{\partial \bm{p}(\bm{r},t)}{\partial t} + \bm{v}(\bm{r},t)\cdot\bm{\nabla}\bm{p}(\bm{r},t)
    		&=    -\frac{d-1}{d\gamma_rc(\bm{r},t)}\frac{\delta F}{\delta\bm{p}}
    			+\left[ \frac{Bd}{d+2}\bm{D}(\bm{r},t) - \bm{\Omega}(\bm{r},t) \right]
    				\cdot\bm{p}(\bm{r},t),\label{eq:polar_dynamics}
\end{align}
\end{subequations}
where $F$ is given in Eq.~(\ref{eq:coarse_grained_fe}).
Inserting these equations into the free energy dissipation and taking functional derivatives
with respect to $\bm{v}$ and $\bm{v}_1$ (with $\bm{v}=(1+\phi)\bm{v}_1/2+(1-\phi)\bm{v}_2/2$) we find the Stokes equation
\begin{equation}\label{eq:NS_final}
    0= \eta\nabla^2\bm{v} - \bm{\nabla}P -(1+\phi)\bm{\nabla}\frac{\delta F}{\delta \phi} - c\bm{\nabla}\frac{\delta F}{\delta c} 
    - \bm{p}\cdot\left(\bm{\nabla}\frac{\delta F}{\delta \bm{p}}\right)^T 
      + \frac{Bd}{2(d+2)}\bm{\nabla}\cdot \left( \bm{p}\frac{\delta F}{\delta \bm{p}} + \frac{\delta F}{\delta \bm{p}}\bm{p} \right)
       -\frac{1}{2}\bm{\nabla}\cdot \left(\bm{p}\frac{\delta F}{\delta \bm{p}} - \frac{\delta F}{\delta \bm{p}}\bm{p} \right)
\end{equation}
where $P$ is the pressure enforcing incompressibility $\bm{\nabla}\cdot\bm{v}=0$.
Note that the $-\bm{\nabla}\delta F/\delta\phi$ term in the expression above can be absorbed into the pressure term $-\bm{\nabla}P$. 
Eqs.~(\ref{eq:dynamics}) and (\ref{eq:NS_final}) completely describe the dynamics of the surfactant in binary fluid system for the case of non-interacting surfactant molecules (they are only interacting through the fluid velocity). 
The mathematical structure of the stress-like tensor in the Stokes equation (\ref{eq:NS_final}) is consistent with the Cauchy stress previously obtained for polar liquid crystals~\cite{cates2018theories,Markovich_2019}.

In the absence of external driving, minimizing the Rayleighian dissipation functional yields the equilibrium equations of motion for the system. When a flow field is imposed, e.g. a shear flow, the Rayleighian simply acquires an additional contribution associated with the externally driven fluid flow. 
Minimizing this modified functional produces the full equations governing both the transient dynamics and the resulting non-equilibrium steady state. 
Thus, the Rayleighian formalism provides a unified description of equilibrium and driven non-equilibrium behaviour, without the need for separate modelling assumptions.

%%%
\section{Results and Discussion\label{sec:results}}
%%%

To demonstrate the validity and reliability of our model, we present three case studies. 
The first examines a planar interface under the assumption of weak coupling, where we show that the system equations can be solved analytically via perturbation theory to any desired order. 
Numerical results are also provided, showing excellent agreement with our analytical solutions. 
The second case study extends this system by using the perturbative solutions to calculate the surface tension of a surfactant-loaded interface, where the presence of surfactants leads to a decrease in surface tension. 
Finally, we demonstrate how the inclusion of surfactants may suppress the coalescence of oil droplets. 

%%%
\subsection{Non-dimensionalization and numerical method of solution\label{sec:num-scheme}}
%%%

All simulation results below are presented in dimensionless units.
Without loss of generality, we set $\alpha = -\beta$ in the free energy expression [Eq.~(\ref{eq:coarse_grained_fe})], so that the bulk free energy favors phase separation into $\phi = \pm 1$.
We choose the interfacial width of the pure binary fluid, $\xi_{\text{I}} = \sqrt{2\kappa/\beta}$ as the unit of length, $\tau$ as the unit of time, and $k_\mathrm{B}T$ as the unit of energy.
We also introduce a small (dimensionless) parameter $\varepsilon = \chi \ell/\left( \xi_{\text{I}} k_\mathrm{B}T \right)$ to represent the weak coupling between the fluid and surfactants. 
Physically $\varepsilon$ characterizes the interaction strength between the surfactant molecules and the binary fluid interface (relative to $k_{\text{B}}T$).
For simplicity, we retain the same notation for the non-dimensionalized constants and variables as in the dimensional forms. 
More details can be found in the Appendix~\ref{app:non-dim}. 
The dimensionless Helmholtz free energy is then given by
\begin{equation} \label{eq:F-dimensionless}
F[\phi,c,\boldsymbol{p}] = \int \left\{ -\frac{\beta}{2}\phi^2 + \frac{\beta}{4} \phi^4 + \frac{\beta}{4}|\boldsymbol{\nabla}\phi|^2 + c\ln(c) 
+\frac{d}{2}c|\boldsymbol{p}|^2 + \varepsilon c\boldsymbol{p}\cdot\boldsymbol{\nabla}\phi \right\} d^dr,
\end{equation}
and the resulting system of coupled partial differential equations governing the dynamics of the binary fluid volume fraction, surfactant concentration, average orientation field, and fluid velocity can be expressed as:
\begin{subequations}
\label{eq:system-eqs-non-dim}
\begin{align}
 \frac{\partial \phi}{\partial t}+(\bm{v}\cdot\bm{\nabla}) \phi = M\nabla^2\left(\frac{\delta F}{\delta\phi}\right) &= M \nabla^2
    \left[-\beta\phi + \beta\phi^3 - \frac{\beta}{2}\nabla^2\phi 
    - \varepsilon\bm{\nabla}\cdot(c\bm{p}) 
    \right], \label{eq:phidot-nondim}\\
\frac{\partial c}{\partial t}+(\bm{v}\cdot\bm{\nabla})c 
       = \boldsymbol{\nabla}\cdot\left[\frac{c}{\gamma_t}\boldsymbol{\nabla} \left(\frac{\delta F}{\delta c}\right) \right] &= 
\frac{1}{\gamma_t}\nabla^2c + \frac{1}{\gamma_t}
\bm{\nabla} \cdot
    \left[ c\bm{\nabla}\left(\frac{d}{2}|\bm{p}|^2+\varepsilon\bm{p}\cdot\bm{\nabla} \phi\right)\right], \label{eq:cdot-nondim} \\
\frac{\partial \bm{p}}{\partial t}+(\bm{v} \cdot \bm{\nabla})\bm{p} + \bm{\Omega}\cdot\bm{p} - \frac{Bd}{d+2}\bm{D}\cdot\bm{p} = 
     -\frac{d-1}{d\gamma_r c} \left(\frac{\delta F}{\delta\boldsymbol{p}}\right) &=  
    -\frac{d-1}{d\gamma_r}(d\bm{p} + \varepsilon\bm{\nabla}\phi), \\
 0 &=-\bm{\nabla}P + \eta\nabla^2\bm{v}+\bm{f} + \bm{\nabla}\cdot\bm{\sigma}, \label{eq:stokes} \\
 0 &= \bm{\nabla} \cdot \bm{v},
\end{align}    
\end{subequations}
where we have defined the body force
\begin{equation}
    \bm{f} \equiv -\phi\bm{\nabla}\frac{\delta F}{\delta \phi}
            -c\bm{\nabla}\frac{\delta F}{\delta c}
            -\bm{p}\cdot\bigg(\bm{\nabla}\frac{\delta F}{\delta \bm{p}}\bigg)^T
\end{equation}
and the stress-like tensor
\begin{equation}
    \bm{\sigma} \equiv \frac{Bd}{2(d+2)}
        \left(\bm{p}\frac{\delta F}{\delta\bm{p}} + 
              \frac{\delta F}{\delta\bm{p}}\bm{p}\right)
        -\frac{1}{2}
        \left(\bm{p}\frac{\delta F}{\delta\bm{p}}
             -\frac{\delta F}{\delta\bm{p}}\bm{p}\right).
\end{equation}
In the dimensionless unit, the bare surface tension is then given by $\sigma_\text{I}=2\beta/3$.
For the remainder of this paper, we consider the two-dimensional case ($d=2$) only.

In the following, we briefly outline the general numerical scheme employed in our study, based on the dimensionless equations~\eqref{eq:system-eqs-non-dim}.
We consider a two-dimensional domain of size $L_x$ in the $x$-direction and $L_y$ in the $y$-direction.
We employ a simple Finite Difference Method with Euler time-stepping to simulate the variables $\phi(\bm{r},t)$, $c(\bm{r},t)$, and $\bm{p}(\bm{r},t)$. 
This is implemented using the Python library NumPy~\cite{harris2020array}. 
Central differences are used for both first- and second-order derivatives.
We assume periodic boundary conditions at $y=0$ and $y=L_y$.
Boundary conditions at $x=0$ and $x=L_x$ can either be no-flux or periodic.
No-flux conditions are derived to conserve $\phi(\bm{r},t)$ and $c(\bm{r},t)$, resulting in either Neumann or reflective boundary conditions, as detailed in Appendix~\ref{app:BC}. 
Unless otherwise specified, the values for all parameters used in the simulations are provided in Appendix~\ref{app:non-dim}. 

At each time step, the fluid velocity $\bm{v}$ is calculated using a pseudospectral method as outlined in Ref.~\cite{hardy2024hybrid}. 
By transforming the Stokes equation~\eqref{eq:stokes} into Fourier space (where Fourier-space variables are denoted with a tilde), we obtain the expression used to calculate the velocity at each time step
\begin{align}
\widetilde{\bm{v}}_{\bm{k}} = \frac{1}{\eta k^2} 
\left[\widetilde{\bm{f}}_{\bm{k}} - \left( \widetilde{\bm{f}}_{\bm{k}}\cdot\hat{\bm{k}} \right) \hat{\bm{k}} \right],  
\end{align}
where $\bm{k}$ is the wavevector, $k = |\bm{k}|$ is the corresponding wavenumber, and $\hat{\bm{k}} = \bm{k}/k$ is the unit wavevector. 
The forcing term $\widetilde{\bm{f}}_{\bm{k}}$ encompasses all terms from $\bm{f}$ and $\bm{\nabla} \cdot \bm{\sigma}$. The Fourier-space pressure is
\begin{align}
    \widetilde{P}_{\bm{k}} = -\frac{i}{k} \widetilde{\bm{f}}_{\bm{k}}\cdot\hat{\bm{k}}.
\end{align}
% This form is derived to ensure the incompressibility condition $\bm{\nabla} \cdot \bm{u} = 0$. 
As with $\phi$, $c$, and $\boldsymbol{p}$, we impose periodic boundary conditions on $\boldsymbol{v}$ at $y=0$ and $y=L_y$.
Along the $x$-direction, the fluid velocity $\boldsymbol{v}$ can satisfy either no-slip or periodic boundary conditions.
In the case of no-slip boundary conditions at $x=0$ and $x=L_x$, combined with periodic boundary conditions at $y=0$ and $y=L_y$, 
it is convenient to use a sine transform in $x$ and a standard Fourier transform in $y$, i.e,
\begin{equation}
\boldsymbol{v}(\boldsymbol{r}) = \sqrt{\frac{2}{L_xL_y}} \sum_{n,m} \widetilde{\boldsymbol{v}}_{\boldsymbol{k}} \sin\left(\frac{n\pi x }{L_x}\right)
		e^{i\frac{2\pi my}{L_y}} ,
\end{equation}
where $n=1,2,3,\dots$ and $m\in\mathbb{Z}$.
The corresponding discrete wavevector $\boldsymbol{k}=(k_x,k_y)^T$ are given by $k_x=\pi n/L_x$ and $k_y=2\pi m/L_y$, see~\cite{code} for detailed implementation.

We validated our model by simulating a pure binary fluid and obtaining the classical solution for a clean binary fluid~\cite{cates2018theories}. 
The pseudospectral scheme used for velocity was also validated through the Poiseuille flow example, where it perfectly matched the analytical solutions.

%%%
\subsection{Planar Interface\label{sec:plane}}
%%%

Our first case study involves a flat, vertical water-oil interface at $x=0$, as shown in Fig.~\ref{fig:surface-tension-static}. 
Over time, surfactants adsorb onto the interface, resulting in a peak in concentration $c(\bm{r},t)$, although some surfactant concentration remains in the bulk phases. 
Additionally, the surfactant molecules align perpendicular to the interface, with the `head' in the water phase and the `tail' in the oil phase, as indicated by the non-zero $x$-component of $\bm{p}(\bm{r},t)$. When $|\bm{p}| > 0$, the molecules are aligned, as opposed to being randomly arranged when $\bm{p} \approx \bm{0}$. 
At equilibrium, the fluid velocity falls to zero, and the system reaches a static configuration. 
Assuming $\varepsilon = \chi \ell / \left(\xi_\text{I} k_\mathrm{B}T \right) \ll 1$, we can solve the steady state solution perturbatively. 

We now solve the system of nonlinear ordinary differential equations \eqref{eq:system-eqs-non-dim} (with $d=2$) using a regular perturbation approach. 
Due to translational symmetry in the $y$-direction, the solutions depend only on the $x$-direction. 
At steady-state equilibrium, the system equations then simplify to
\begin{subequations}
	\begin{align}
		\beta \left( \phi - \phi^3 + \tfrac{1}{2} \, \phi'' \right) + \varepsilon \, (cp_x)' &= 0 \, , \label{eq:1} \\
		c' + c (p_x^2)' + \varepsilon c (p_x \phi')' &= 0 \, , \label{eq:2} \\
		2p_x + \varepsilon \phi' &= 0 \, , \label{eq:3}
	\end{align}
\end{subequations}
where $p_x$ is the $x$-component of the field $\bm{p}(\bm{r},t)$. 
Here, primes indicate differentiation with respect to $x$. 
It follows from Eq.~\eqref{eq:3} that 
\begin{equation}
	p_x = -\tfrac{1}{2} \, \varepsilon \phi' \, . 
	\label{eq:p}
\end{equation}
Thus, we only need to find the solution for the fields \( \phi \) and \( c \). 
Substituting Eq.~\eqref{eq:p} into Eqs.~\eqref{eq:1} and~\eqref{eq:2}, we obtain
\begin{subequations}
	\begin{align}
		\beta \left( 2\phi - 2\phi^3 + \phi'' \right) - \varepsilon^2 \, (c\phi')' &= 0 \, , \label{eq:1_withoutP} \\
		2c' +  \varepsilon^2 c \phi' \phi'' -\varepsilon^2 c ( {\phi'}^2)' &= 0 \, . \label{eq:2_withoutP}
	\end{align} \label{eq:main}
\end{subequations}
Eqs.~\eqref{eq:main} are highly nonlinear, making an analytical solution challenging. 
To address this, we seek an approximate solution using the perturbation method, expressing the solution as
\begin{subequations}
	\begin{align}
		\phi(x) &= \phi_0(x) + \phi_2(x) \, \varepsilon^2 + \phi_4(x) \, \varepsilon^4 + \mathcal{O} \left( \varepsilon^6 \right) , \label{eq:phi-perturbative} \\
		c(x) &= c_0(x) + c_2(x) \, \varepsilon^2 + c_4(x) \, \varepsilon^4 + \mathcal{O} \left( \varepsilon^6 \right) . \label{eq:c-perturbative} 
	\end{align}
\end{subequations}
Due to parity considerations, we expect the odd powers in the series expansions of \( \phi \) and \( c \) to vanish. Inserting the perturbation expansion into Eqs.~\eqref{eq:main} yields the zeroth-order problem as
\begin{subequations}
	\begin{align}
		\phi_0'' + 2\phi_0 - 2\phi_0^3 &= 0 \, , \\
		c_0' &= 0 \, ,
	\end{align}
\end{subequations}
the solution of which is given by
\begin{subequations}
	\begin{align}
		\phi_0(x) &= \tanh x \, , \\ 
		c_0(x) &= C_0 = \text{const.} \, ,
	\end{align}
\end{subequations}
where \(\tanh\) represents the hyperbolic tangent function and $C_0$ is some constant. To simplify the equations, we introduce the abbreviation \( \lambda = C_0/\beta \) as well as the function \( S(x) = \operatorname{sech}^2 x \).
The order \( \varepsilon^2 \) equations are obtained as
\begin{subequations}
	\begin{align}
		\beta \phi_2'' + 2\beta \left( 1-3\phi_0^2 \right) \phi_2 - C_0 \phi_0''&= 0 \, , \\
		2 c_2' - C_0 \phi_0' \phi_0'' &= 0 \, .
	\end{align}
\end{subequations}
The solution to this is given by
\begin{subequations}
	\begin{align}
		\phi_2(x) &= \tfrac{1}{2} \, \lambda x S(x) \, , \\[3pt]
		c_2(x) &= \tfrac{1}{4} \, C_0 S(x)^2 \, .
	\end{align}
\end{subequations}
The equations at order \( \varepsilon^4 \) are given by
\begin{subequations}
	\begin{align}
		\beta \phi_4'' + 2\beta ( 1- 3\phi_0^2 ) \phi_4 &= c_2\phi_0''+C_0\phi_2'' + c_2'\phi_0' + 6\beta \phi_0\phi_2^2 \, ,  \\[3pt]
		2c_4'  &= C_0 \phi_0' \phi_2'' + \left( C_0 \phi_2'+c_2 \phi_0' \right) \phi_0'' \, .
	\end{align}
\end{subequations}
The solution to this is given by
\begin{subequations}
	\begin{align}
		\phi_4(x) &= \tfrac{1}{16} \, \lambda S(x)
		\left( 6\lambda x + \left( 2 -4\lambda x^2 +  S(x) \right)\tanh x \right)  \, , \\[3pt]
		c_4(x) &= \tfrac{1}{32} \, C_0 S(x)^2 
		\left( 8\lambda \left( 1-2 x \tanh x\right) +  S(x)^2 \right) .
	\end{align}
\end{subequations}
In theory, we could extend beyond the current order and determine higher-order terms in the perturbation series. 
However, due to technical limitations and the complexity of the resulting expressions, we stop at the current order. We expect the series to provide an accurate approximation of the solution, provided that \( \varepsilon \ll 1 \).

From Eq.~\eqref{eq:p}, it follows that the polarization can be expressed as an odd power series in \( \varepsilon \), given by
\begin{align}
	p_x(x) = p_{x1}(x) \varepsilon + p_{x3}(x) \varepsilon^3 + p_{x5}(x) \varepsilon^5 + \mathcal{O} \left( \varepsilon^7 \right) \, . \label{eq:p-perturbative} 
\end{align}
We find after calculation that,
\begin{subequations}
	\begin{align}
		p_{x1}(x) &= -\tfrac{1}{2} \, S(x) \, , \\[3pt]
		p_{x3}(x) &= \tfrac{1}{4} \, \lambda S(x) \left( 2x\tanh x-1\right) \, , \\[3pt]
		p_{x5}(x) &= \tfrac{1}{32} \, \lambda S(x)
		\big( 4-6\lambda-8\lambda x^2 +20\lambda x \tanh x  +\left( 12\lambda x^2-2\right) S(x)
		-5 S(x)^2 \big) \, .
	\end{align}
\end{subequations}

\begin{figure}[t]
  \begin{subfigure}[t]{\textwidth}
    \includegraphics[width=0.65\linewidth]{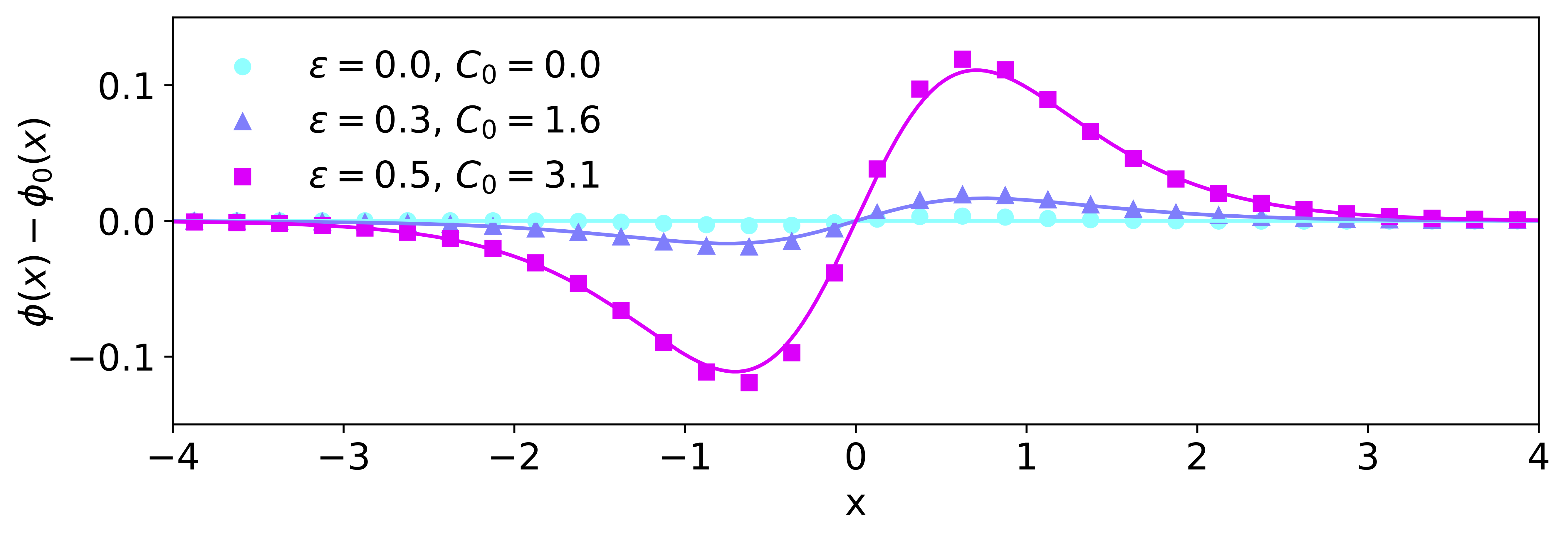}
    \caption{}
    \label{fig:interface-phi}
  \end{subfigure}

  \vspace{0.5em}

  \begin{subfigure}[t]{\textwidth}
    \includegraphics[width=0.65\linewidth]{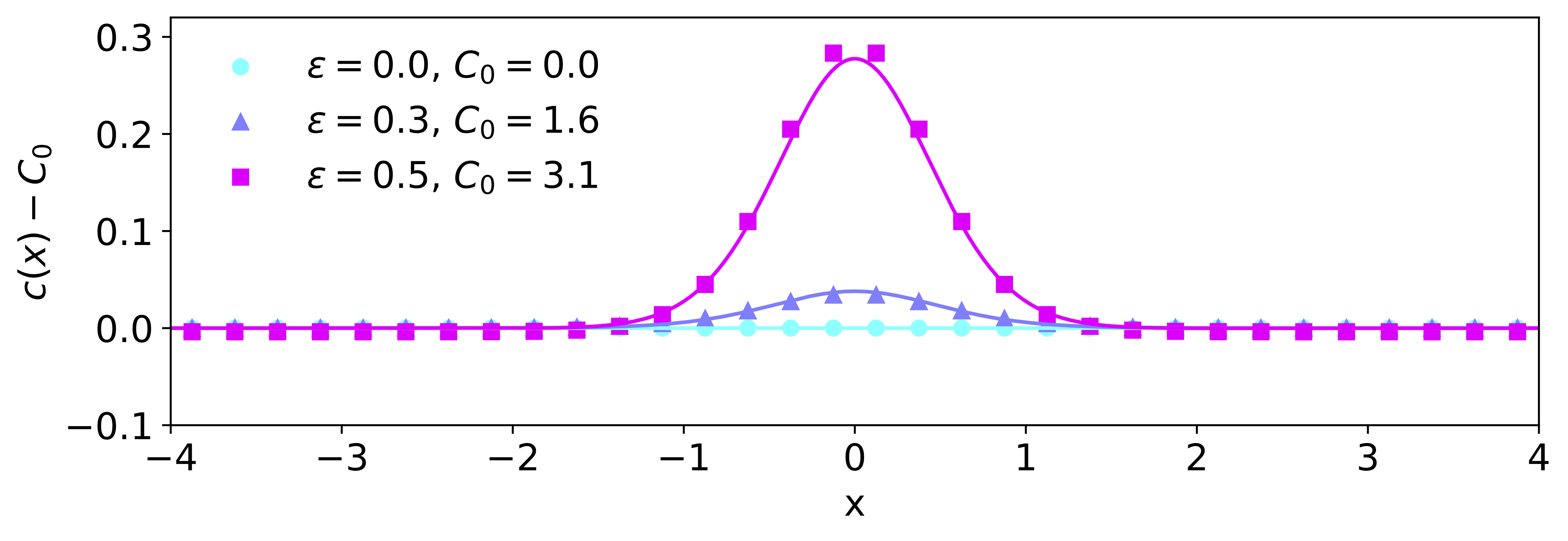}
    \caption{}
    \label{fig:interface-c}
  \end{subfigure}

  \vspace{0.5em}

  \begin{subfigure}[t]{\textwidth}
    \includegraphics[width=0.65\linewidth]{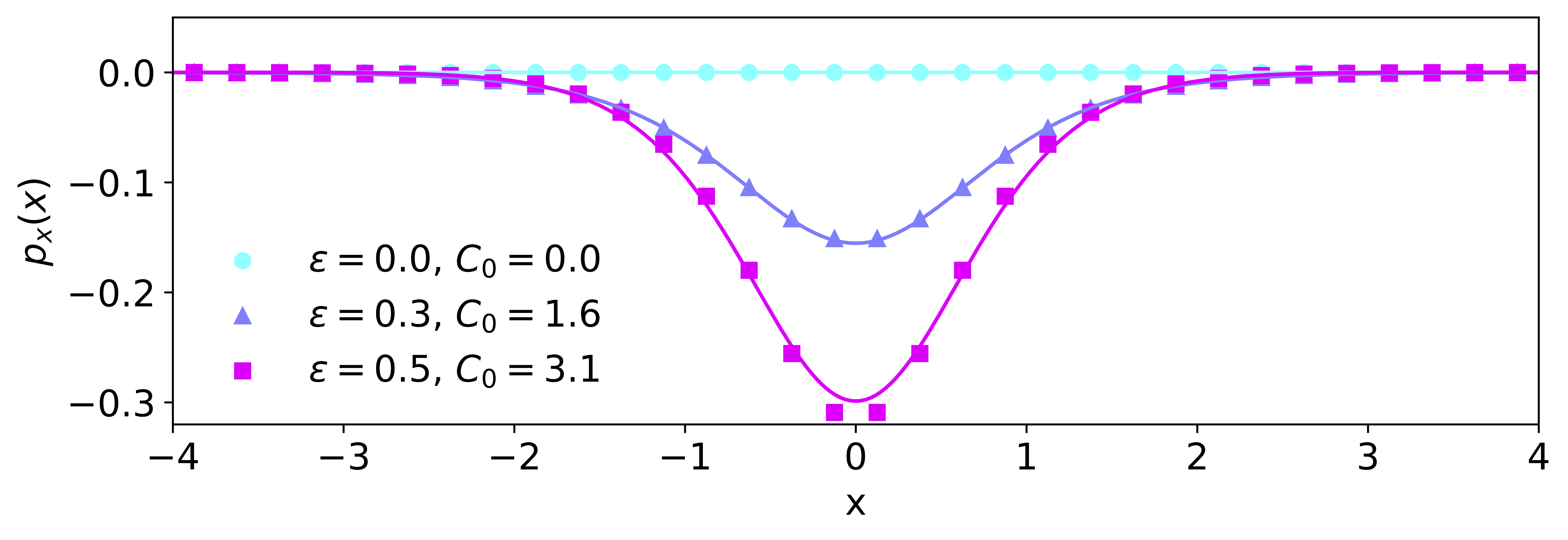}
    \caption{}
    \label{fig:interface-p}
  \end{subfigure}
  
    \caption{\justifying
    (a) A graph showing the analytical (line) and numerical (symbols) solutions for the fluid field $\phi(x)$ with the leading order $\phi_0(x) = \tanh{x}$ removed, at equilibrium for a variety of $\varepsilon$ and $C_0$ values.
    (b) A graph showing the analytical (line) and numerical (symbols) solutions for concentration, $c(x)$ with the leading order $c_0(x)=C_0$ removed, at equilibrium for a range of $\varepsilon$ and $C_0$ values.
    (c) A graph showing the analytical (line) and numerical (symbols) solutions for the polarization field $p_x(x,t)$, at equilibrium for a range of $\varepsilon$ and $C_0$ values. Parameters used: $\beta=2.0$, $B=0.5$, $M=3$, $\gamma_t=\gamma_r=0.01$.}
  \label{fig:interface}
\end{figure}

For comparison, we also simulate the same scenario using the numerical scheme described in Sec.~\ref{sec:num-scheme}. 
We use a two-dimensional rectangular computational domain sized $L_x=64$  and $L_y=4$ with a spatial resolution of $\Delta x=\Delta y=0.25$ and temporal resolution of $\Delta t=0.0001$. 
We use a uniform initial condition for $c(\bm{r},t)$, with initial concentration value $c(\bm{r},t = 0) = C_0$.
Physically, $C_0$ represents the average surfactant concentration, which is conserved throughout.
Volume fraction field $\phi(\bm{r},t)$ is initialized as two halves, with $\phi(x < 0) = -1.0$  (water) and $\phi(x > 0) = 1.0$ (oil). 
Each phase has a magnitude of $1$ as that is the solution of the double well potential in the free energy density. 
All other fields are initialized as zero everywhere. 
We have periodic boundary conditions at $y=0$ and $y=L_y$, and no-slip/no-flux boundary conditions at $x=-L_x/2$ and $x=L_x/2$. 
This is to ensure that only one water-oil interface is created. 
We choose to vary two parameters, $\varepsilon$ and the average surfactant concentration $C_0$. 
Three simulations were run in total, with parameters: $\varepsilon = 0.0$ and $C_0 = 0$ (clean system), one with $\varepsilon = 0.3$ and $C_0 = 1.6$, and another with $\varepsilon = 0.5$ and $C_0 = 3.1$. 

Figs.~\ref{fig:interface} show the steady-state profiles $\phi(x,t\rightarrow\infty)$, $c(x,t\rightarrow\infty)$, and $p_x(x,t\rightarrow\infty)$, respectively, demonstrating excellent agreement with the perturbation theory. 
The deviations in $\phi(\bm{r},t)$ remain small relative to the bulk values ($\pm 1$), indicating that the fluid is only weakly perturbed, consistent with earlier studies, which also assume weak coupling between the fluid density and surfactant concentrations~\cite{van2006diffuse,liu2010phase}, although these studies did not incorporate an explicit polarization field $\bm{p}(\bm{r},t)$. 
As expected, the magnitude of deviations increases with both the coupling strength $\varepsilon$ and the mean surfactant concentration $C_0$. 
The concentration profile $c(\bm{r},t\rightarrow\infty)$ exhibits a peak localized around the interface, with enhancements on the order of $0.2$, comparable to values reported in prior numerical studies~\cite{van2006diffuse,zong2020modeling,liu2010phase}, again in the absence of a polarization field.

As $\varepsilon$ increases, more surfactant molecules are adsorbed at the interface due to the stronger attraction, leading to a higher peak in the surfactant concentration. 
The absorbed molecules exhibit alignment, as indicated by the peak in $p_x(x,t\rightarrow\infty)$ near the interface. 
Since the leading-order contribution to $p_x(x,t\rightarrow\infty)$ is zero, this peak arises entirely from perturbative corrections. 
At equilibrium, molecular orientation at the interface exhibits minimal noise, with the $y$-component of $\bm{p}(\bm{r},t\rightarrow\infty)$ remaining negligible. 
We observe little dependence of the alignment strength on $C_0$, indicating that increasing the number of adsorbed surfactant molecules does not significantly affect their collective orientation. 
Instead, alignment is more strongly governed by the coupling strength $\varepsilon$, which determines the effective attraction to the interface. 
Overall, the simulations show excellent agreement with the perturbative theory and are consistent with established understanding of surfactant behavior at fluid interfaces at microscopic level.

%%%
\subsection{Adsorption Isotherm and Effective Surface Tension}
%%%

\begin{figure}

  \begin{subfigure}{0.305\textwidth}
      \includegraphics[width=5.7cm]{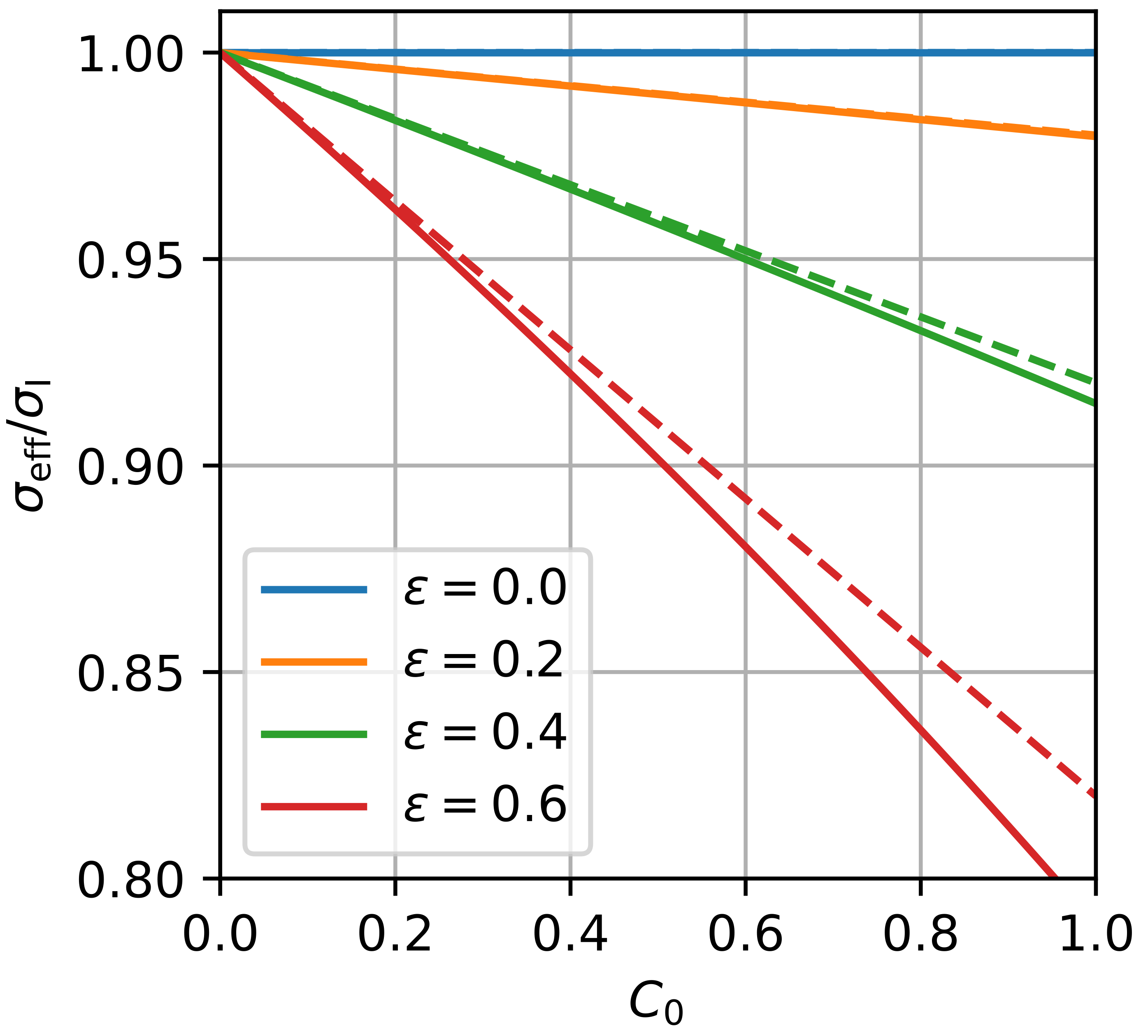}
      \caption{}
    \label{fig:surface-tension-plot}
  \end{subfigure}
  \begin{subfigure}{0.305\textwidth}
      \includegraphics[height=5.7cm]{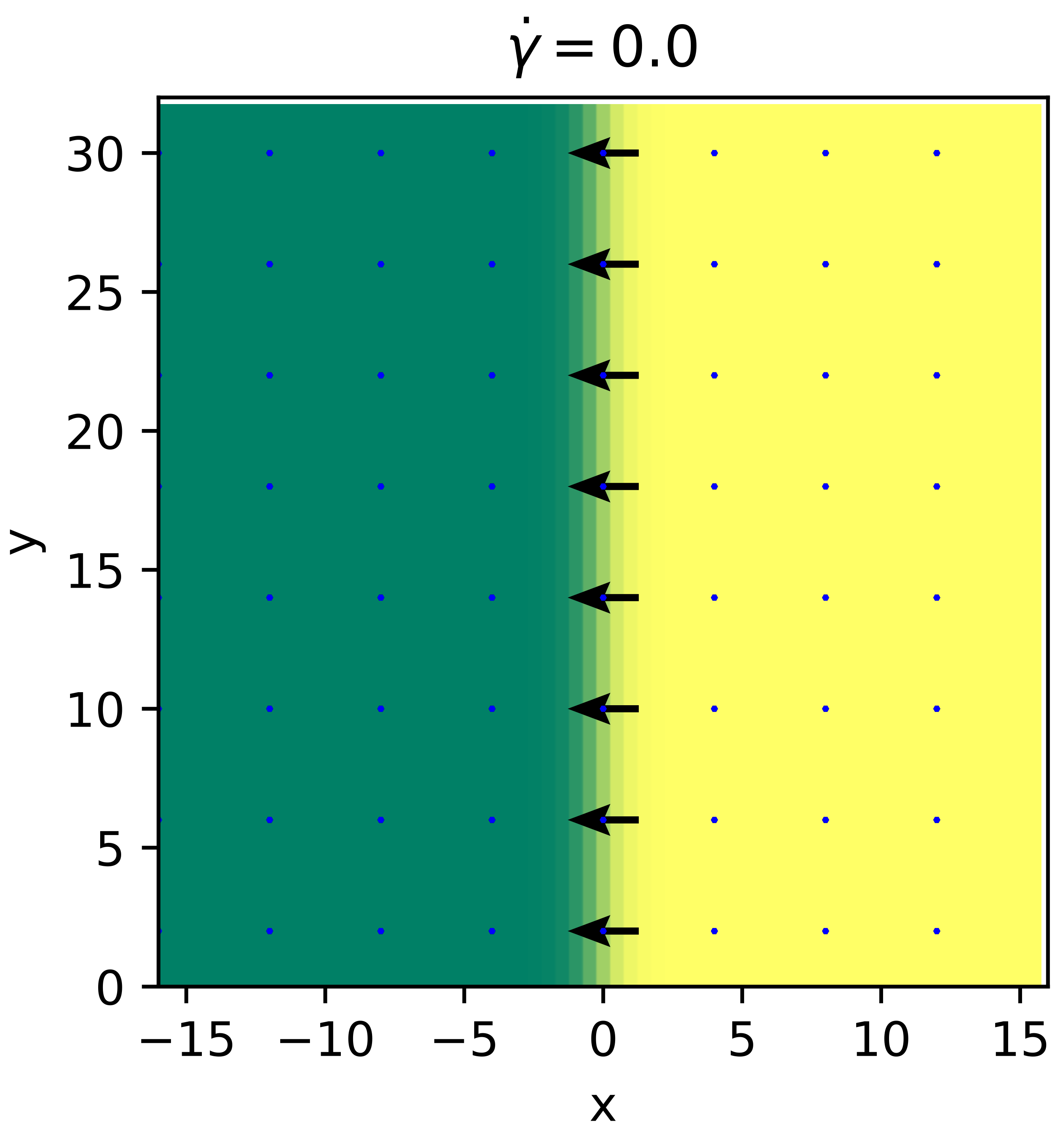}
      \caption{}
    \label{fig:surface-tension-static}
  \end{subfigure}
  \begin{subfigure}{0.375\textwidth}
      \includegraphics[height=5.7cm]{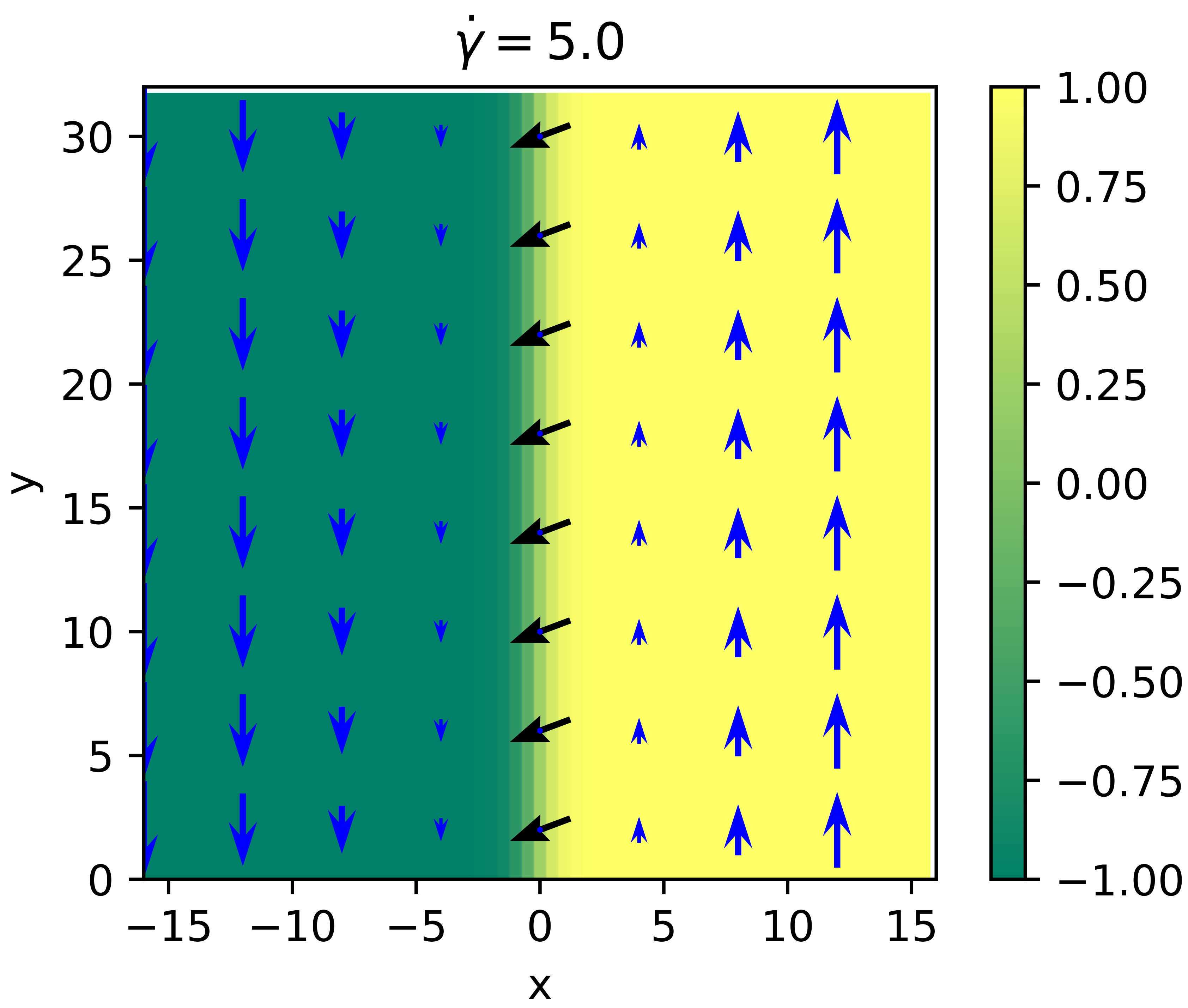}
      \caption{}
    \label{fig:surface-tension-shear}
  \end{subfigure}

  \caption{\justifying
  (a) Effective surface tension divided by the bare surface tension $\sigma_{\text{eff}}/\sigma{\text{I}}$ as a function of bulk surfactant concentration $C_0$ for different values of coupling strength $\varepsilon$ and fixed $\beta=1$. Dashed lines indicate the leading-order prediction from the Gibbs isotherm.
    (b) Equilibrium configuration of a planar interface located at $x=0$. Black arrows show the polarization field $\boldsymbol{p}$ which aligns perpendicular to the interface.
    (c) Under strong shear flow, the polarization $\boldsymbol{p}$ field becomes tilted and is no longer perpendicular to the interface. Blue arrows indicate the fluid velocity $\boldsymbol{v}$.
    Parameters used: $\beta=1.0$,  $B=1.0$, $M=1.0$, $\gamma_t=\gamma_r=0.1$, $\eta=1.0$, and $\varepsilon=0.5$.}
    \label{fig:surface-tension}
\end{figure}

Let us consider a flat interface at $x=0$ with $\phi(x<0)<0$ (water phase) and $\phi(x>0)>0$ (oil phase), as shown in Fig.~\ref{fig:surface-tension-static}.
The concentration profile $c(x)$ across this interface has been solved perturbatively in Eq.~(\ref{eq:c-perturbative}) and plotted in Fig.~\ref{fig:interface-c}.
In experiments, the volumetric concentration of surfactant at an oil–water interface is typically $10^2$-$10^4$ times larger than the bulk concentration~\cite{Vishnyakov_2022}, far exceeding the interfacial concentrations reached in our simulations.
This occurs because real fluid interfaces are extremely sharp: the interfacial width is on the order of a molecular length scale.
By contrast, in simulations we employ a diffuse-interface (phase-field) model, in which the interface is spread over a finite width $\xi_{\text{I}}$.
For numerical stability and resolution, $\xi_{\text{I}}$ is usually taken to be several orders of magnitude larger than its physical, molecular-scale value~\cite{Jaensson_2017}.
Instead, we quantify adsorption by measuring the amount of surfactant adsorbed per unit interfacial area,
\begin{equation}
\Gamma = \int_{-\infty}^\infty \left[ c(x) - C_0 \right] dx.
\end{equation}
Substituting the perturbative solution for $c(x)$ in Eq.~(\ref{eq:c-perturbative}) to the above expression, we obtain
\begin{equation}
\Gamma = \frac{1}{3}\epsilon^2 C_0 + \mathcal{O}(\epsilon^4). \label{eq:Gamma}
\end{equation}
Thus, the amount of adsorbed surfactant increase linearly with the bulk concentration $C_0$. 
This is consistent with the Henry's law adsorption isotherm in the dilute surfactant limit~\cite{van2006diffuse}.
A full quantitative comparison with adsorption–desorption kinetics would require extending the model to include finite-rate kinetic terms that explicitly describe the adsorption and desorption of molecules at the interface.

We can now compute the reduction in effective surface tension $\sigma_{\text{eff}}$ induced by the adsorption of surfactant molecules at the interface.
To do so, we first define the grand potential functional to be:
\begin{equation}
\Phi[\phi,c,\boldsymbol{p}] = F[\phi,c,\boldsymbol{p}] - \bar{\mu}_c\int c(\boldsymbol{r})\,d^dr \equiv \int w(\mathbf{r}) \,d^dr,
\end{equation}
where $w(\boldsymbol{r})$ denotes the grand potential density.
The constant $\bar{\mu}_c$ is the chemical potential of an external surfactant reservoir, corresponding to the grand canonical ensemble.
Mathematically, the grand potential $\Phi[\phi,c,\boldsymbol{p}]$ is a Legendre transform of the Helmholtz free energy $F[\phi,c,\boldsymbol{p}]$ with respect to $c$.
In equilibrium, the condition $\delta F/\delta c=\bar{\mu}_c$ yields $\bar{\mu}_c=\log(C_0)+1$.
The effective surface tension $\sigma_{\text{eff}}$ is then defined to be the excess grand potential per unit area of the interface:
\begin{equation}
\sigma_{\text{eff}} = \int_{-\infty}^{\infty} \left[ w(x) - w_{\text{bulk}} \right] dx, \label{eq:sigma-eff}
\end{equation}
where $w_{\text{bulk}}$ is the value of the grand potential density evaluated in the bulk, i.e. in the limit $x\rightarrow\pm\infty$, or far from the interface.
Using the perturbative solutions for $\phi(x)$, $c(x)$, and $p_x(x)$ from Sec.~\ref{sec:plane}, the integral above can then be computed numerically, retaining terms up to order $\mathcal{O}(\varepsilon^6)$.
The resulting effective surface tension $\sigma_{\text{eff}}$ as a function of the bulk surfactant concentration $C_0$ is shown in Fig.~\ref{fig:surface-tension-plot} (solid lines). 
As expected, $\sigma_{\text{eff}}$ approaches the bare interfacial value $\sigma_{\text{I}}=2\beta/3$ of the clean binary fluid in the limit $C_0\rightarrow\infty$.
For comparison, we may also invoke the Gibbs adsorption isotherm, which in dimensionless form reads:
\begin{equation}
\frac{d\sigma_{\text{eff}}}{d\ln C_0} = -\Gamma.
\end{equation}
Substituting the expression for $\Gamma$ from Eq.~(\ref{eq:Gamma}) and integrating, we obtain the explicit leading-order result for the effective surface tension
\begin{equation}
 \sigma_{\text{eff}} = \sigma_{\text{I}} - \frac{\varepsilon^2}{3}C_0 + \mathcal{O}(\varepsilon^4).
\end{equation}
The corresponding prediction from the Gibbs isotherm is shown in Fig.~\ref{fig:surface-tension-plot} (dashed lines), where it agrees with our direct definition of $\sigma_{\text{eff}}$ in Eq.~(\ref{eq:sigma-eff}).

Finally, we note that the effective surface tension shown in Fig.~\ref{fig:surface-tension-plot}  applies only under equilibrium conditions.
At equilibrium, the fluid velocity vanishes $\boldsymbol{v}=\boldsymbol{0}$, and the fluid composition $\phi$, surfactant concentration $c$ and polarization $\boldsymbol{p}$ relax to the configuration which minimizes the free energy in Eq.~(\ref{eq:F-dimensionless}).
In this state, the polarization field $\boldsymbol{p}$ aligns strictly perpendicular to the interface, see Fig.~\ref{fig:surface-tension-static}.
In contrast, when the system is subjected to a strong shear flow, for example $\boldsymbol{v}\simeq\dot{\gamma}x\hat{\boldsymbol{y}}$,
the polarization $\boldsymbol{p}$ is oriented by the flow and becomes tilted relative to the interface normal vector, see Fig.~\ref{fig:surface-tension-shear}.
This shear-induced reorientation modifies the local interfacial stresses and, consequently, can change the effective surface tension.

\begin{figure*}[t]
    \centering
    \includegraphics[width=0.6\linewidth]{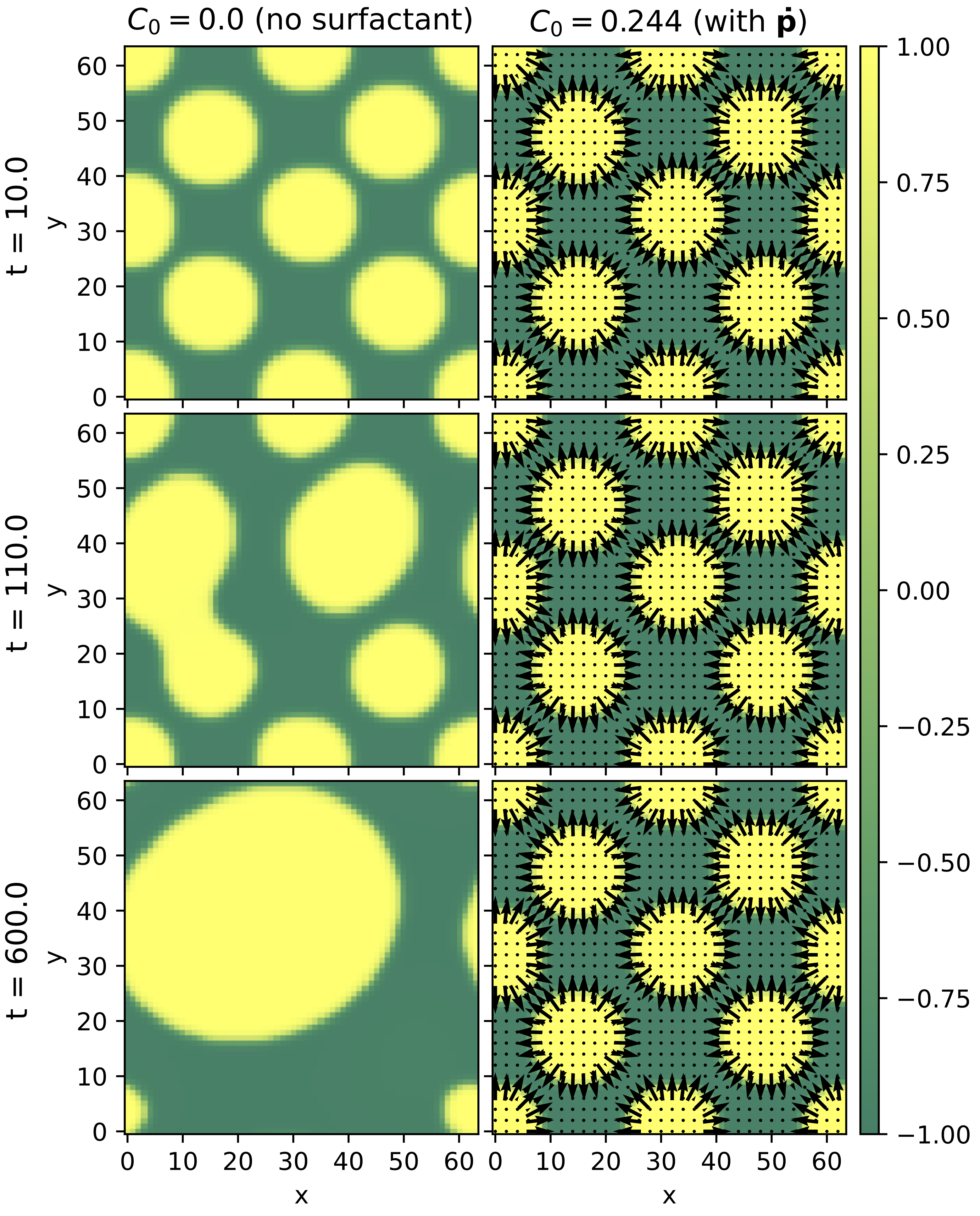}
    \caption{\justifying
    Plots of binary fluid volume fraction $\phi(\bm{r},t)$ for bare emulsion (left column) and surfactant-containing emulsion (right column) at different time steps (rows) with values $t = 10, 110$ and $600$ being the first, second and third rows respectively. 
    The black arrows on the right column indicate the polarization or average orientation of the surfactant molecules, $\bm{p}(\bm{r},t)$.
    The graphs show that the presence of the surfactants suppresses droplet coalescence and full phase separation.
    Parameters used: $(\varepsilon=0,C_0=0)$ (left column) and $(\varepsilon=1.5,C_0=0.244)$ (right column). 
    Other parameters: $\beta=2$, $B=0.5$, $M=3$, and $\gamma_t=\gamma_r=0.01$.}
    \label{fig:emulsion}
\end{figure*}

%%%
\subsection{Emulsion Study}
%%%

Our final case study examines an emulsion of oil droplets suspended in water, with the initial condition shown in the first row of Fig.~\ref{fig:emulsion}. 
Simulations were performed on a $L_x\times L_y=64 \times 64$ grid with spatial resolution $\Delta x = \Delta y = 0.5$ and time step $\Delta t = 0.001$. 
To compare the effects of surfactants, we conducted two simulations: one for a clean system with no surfactants ($\varepsilon = 0$, $C_0 = 0$), and one with surfactants present ($\varepsilon = 1.5$, $C_0 = 0.244$). 
Periodic boundary conditions were applied along all edges of the domain ($x = 0$, $x = L_x$, $y = 0$, $y = L_y$).

In the absence of surfactants, the droplets gradually coalesce into a single large drop. By contrast, when surfactants are present, coalescence is significantly suppressed, leading to a more stable emulsion—although the system will still slowly evolve toward full phase separation over longer timescales. 
This stabilization arises from the polarization field $\bm{p}(\bm{r},t)$: the surfactant molecules orient outward from each droplet, generating effective repulsive interactions that hinder droplet merging. 
As shown in Fig.~\ref{fig:emulsion}, the clean system (left column) exhibits progressive coalescence, whereas the system with surfactants (right column) retains a dispersed droplet morphology over the duration of the simulation. 
Although not shown, we note that increasing the mean surfactant concentration $C_0$ further suppresses coalescence, much like increasing the interaction strength $\varepsilon$.
Finally, as shown in Fig.~\ref{fig:emulsion} (right column), the surfactant orientation vectors $\bm{p}(\bm{r},t)$ point outward from each droplet. 
In this configuration, adjacent droplets effectively experience mutual repulsion due to opposing molecular orientations, thereby maintaining separation and stabilizing the emulsion.

For comparison, we also performed a separate set of simulations in which the polarization field is treated as quasi-static.
In this approximation, $\boldsymbol{p}$ is assumed to relax instantaneously to the configuration that minimizes the free energy, namely
\begin{equation}
\frac{\delta F}{\delta\boldsymbol{p}} = \boldsymbol{0} \quad\Rightarrow\quad \boldsymbol{p} = -\frac{\varepsilon}{2}\boldsymbol{\nabla}\phi.
\end{equation} 
Under weak-flow conditions, where $\boldsymbol{v}\simeq\boldsymbol{0}$, we find no qualitative difference in the stability of the emulsion compared with fully dynamical model shown in Fig.~\ref{fig:emulsion} (right column).
However, in the presence of a strong imposed shear, similar to that in Fig.~\ref{fig:surface-tension-shear}, some quantitative and/or qualitative difference may emerge.

%%%
\section{Conclusion}
%%%

In this work, we have developed a continuum hydrodynamic model for binary-fluid–surfactant systems by systematically deriving both the microscopic and coarse-grained dynamics using Rayleigh’s minimum energy dissipation principle.
At the microscopic level, surfactant molecules are modelled as rigid dumbbells that undergo Brownian motion and exert forces and torques on the surrounding fluid due to their amphiphilic character, while the fluid is treated as a continuum field—resulting in a hybrid discrete–continuum formulation.
By constructing an appropriate Rayleighian dissipation functional, we derived the overdamped stochastic equations governing the translational and rotational dynamics of individual surfactant molecules.

Upon coarse-graining these microscopic dynamics, we derived a closed set of continuum equations for the surfactant concentration field $c(r,t)$, polarization field 
$\bm{p}(\bm{r},t)$, binary fluid composition $\phi(\bm{r},t)$, and fluid velocity $\bm{v}(\bm{r},t)$. 
These equations follow consistently from a single mesoscopic free energy functional, ensuring thermodynamic consistency and preserving detailed balance at equilibrium.

Our model successfully captures key physical phenomena associated cwith surfactants, including their accumulation and alignment at interfaces, the reduction of surface tension, and the suppression of droplet coalescence in emulsions. 
These predictions were validated through a combination of perturbative analytical solutions and direct numerical simulations. 
Unlike previous approaches, our framework requires no ad-hoc stabilizing terms, as all interfacial effects and coupling mechanisms emerge naturally from the underlying variational formulation.

Beyond passive surfactants, our formalism is sufficiently general to accommodate additional microscopic ingredients.
In particular, the Rayleighian framework and the polarization dynamics can be readily extended to describe active particles at fluid-fluid interfaces, where self-propulsion, active forces and torques can modify the interfacial stresses~\cite{saintillan2013active}.
This opens the door to studying a broad class of nonequilibrium interfacial phenomena such as active emulsions, self-assembled interfacial monolayers, and active Marangoni flows.

\begin{acknowledgements}
AJH acknowledges EPSRC DTP Studentship No. 2739112. 
SC and ET acknowledges funding from EPSRC Grant No. EP/W027194/1.
SMD acknowledges London Mathematical Society Undergraduate Research Bursary URB-2024-40.
\end{acknowledgements}

\section*{Declaration of Interests}
The authors report no conflict of interest.

\section*{Use of Generative AI}
The authors acknowledge the use of generative AI during the preparation of this manuscript for the purpose of improving grammar, enhancing clarity, and supporting literature searches. All outputs produced by the tool were carefully checked and revised by the authors, who assume full responsibility for the integrity and accuracy of the published work.

\appendix

%%%
\section{Alternative derivation of the coarse-grained free energy\label{app:alt_deriv}}
%%%

In Sec.~\ref{sec:quiescent} and \ref{sec:coarse-graining}, we derived the free energy functional using the Rayleigh dissipation functional. 
In this section, we show that one can also derive the free energy functional using direct thermodynamic arguments from Shannon entropy. 
The total free energy of the system is
\begin{align}
F_{\text{total}}[\phi,c,\bm{p}] = F_{\text{fluid}}[\phi] + F_{\text{surfactants}}[\phi,c,\bm{p}] \label{eq:free energy sum} \, , 
\end{align}
composed of $F_{\text{fluid}}[\phi]$, the free energy contribution from the binary fluid, and $F_{\text{surfactants}}[\phi,c,\bm{p}]$, the free energy contribution from the surfactant molecules. 
The fluid free energy is taken from standard Cahn-Hilliard theory~\cite{lee2014physical}, while the surfactant contributions are derived using the Helmholtz equation:
\begin{align}
    F_{\text{surfactants}}[\phi,c,\bm{p}] = U[\phi,c,\bm{p}] - TS[c,\bm{p}] \, ,
\end{align}
where $U$ is the potential energy, $T$ is the temperature, and $S$ is the Shannon entropy. 
The potential energy for the interaction between the molecules and the fluid can be calculated by summing the microscopic energy [second term in Eq.~\eqref{eq:free_energy_2}] over all particles $i=1,2,\dots N$, where $N$ is the total number of particles.
\begin{align}
    U[\phi,\{\bm{r}_i,\hat{\bm{e}}_i\}] = \chi \ell \sum_{i=1}^{N}\hat{\bm{e}}_i\cdot\bm{\nabla}_i\phi(\bm{r}_i).
\end{align}
Note that we have ignored higher order terms $\sim\ell^2$.
The equation above can then be coarse-grained straight into its continuum form:
\begin{align}\label{eq:energy_direct}
U[\phi,c,\bm{p}] = \int_V \left(\chi \ell c\bm{p}\cdot\bm{\nabla}\phi\right) dV,
\end{align}
by approximating
\begin{equation}
c(\bm{r},t)\simeq \sum_{i=1}^N \delta(\bm{r}-\bm{r}_i) \quad\text{and}\quad 
c(\bm{r},t)\bm{p}(\bm{r},t) \simeq \sum_{i=1}^N \hat{\bm{e}}_i\delta(\bm{r}-\bm{r}_i).
\end{equation}
The entropic term can be derived using Shannon's information entropy expression together with our probability density $\psi(\bm{r},\hat{\bm{e}},t)$:
\begin{align}
    S[c,\bm{p}] = -k_\mathrm{B}\int d^dr\int d^{d-1}\hat{e} \, \left[\psi\ln(\mathcal{N}\psi)\right]. \label{eq: entropy}
\end{align}
Here, $\mathcal{N}$ is a constant ensuring the argument of the logarithm remains dimensionless. 
This expression can be simplified by representing $\psi(\bm{r},\hat{\bm{e}},t)$ as a spherical harmonics expansion (see Appendix~\ref{app:distribution_expansion}):
\begin{align}
    \psi = \frac{1}{S_d}c + \frac{d}{S_d}c\bm{p}\cdot\hat{\bm{e}} \, .
\end{align}
where $S_d$ is the surface area of a $d$-dimensional unit sphere. Higher-order terms are assumed to be negligible. 
Substituting this into equation~\eqref{eq: entropy} yields:
\begin{align}
    S[c,\bm{p}] &= -\frac{k_\mathrm{B}}{S_d}\int d^dr\int d^{d-1}\hat{e} 
    \left(c + dc\bm{p}\cdot\hat{\bm{e}}\right)\left[\ln\left(\frac{\mathcal{N}c}{S_d}\right)+\ln\left(1+d\bm{p}\cdot\hat{\bm{e}}\right)\right].
\end{align}
Expanding the second logarithmic term for small $\bm{p}$ gives:
\begin{align}
    S[c,\bm{p}] &= -\frac{k_\mathrm{B}}{S_d}\int d^dr\int d^{d-1}\hat{e} 
    \left[c\ln\left(\frac{\mathcal{N}c}{S_d}\right)\left( 1+ d\bm{p}\cdot\hat{\bm{e}}\right)+dc\bm{p}\cdot\hat{\bm{e}} +\frac{1}{2}d^2c\left(\bm{p}\cdot\hat{\bm{e}}\right)^2 \right].
\end{align}
Integrating with respect to $\hat{\bm{e}}$ and using the results from Appendix~\eqref{app:distribution_expansion} give us the final expression for the entropic contribution from the surfactants:
\begin{align}
    S[c,\bm{p}] = -k_\mathrm{B}\int d^dr \left[c\ln\left(\frac{\mathcal{N}}{S_d}c\right)+\frac{d}{2}c|\bm{p}|^2\right]dV \, .
\end{align}
Combining Eq.~\eqref{eq:energy_direct} with this entropy contribution along with the free energy of the fluid $F_{\text{fluid}}[\phi]$ yields the coarse-grained free energy of Eq.~\eqref{eq:coarse_grained_fe}.

%%%
\section{Moment-averaging the single-particle distribution function\label{app:distribution_expansion}}
%%%

We expand the distribution function $\psi(\bm{r}, \hat{\bm{e}}, t)$ in terms of spherical harmonics as follows:
\begin{align}
    \psi(\bm{r}, \hat{\bm{e}}, t)=A_0 c(\bm{r}, t)+A_1 c(\bm{r}, t) \bm{p}(\bm{r}, t) \cdot \hat{\bm{e}}+\cdots.\label{eq: psi expansion}
\end{align}
We assume higher order harmonics to be small so we may set $A_n=0$ for all $n \geq 2$. We are now left with two coefficients $A_0$ and $A_1$, which are yet to be determined. To find these coefficients, we need the following results:
\begin{subequations}
   \begin{align}
\int \hat{e}_\alpha \, d \Omega_d & =0 \, ,\\
\int \hat{e}_\alpha \hat{e}_\beta \, d \Omega_d & =\frac{S_d}{d} \delta_{\alpha \beta} \, , \\
\int \hat{e}_\alpha \hat{e}_\beta \hat{e}_\gamma \, d \Omega_d & =0 \, , \\
\int \hat{e}_\alpha \hat{e}_\beta \hat{e}_\gamma \hat{e}_\delta \, d \Omega_d & =\frac{S_d}{d(d+2)}\left(\delta_{\alpha \beta} \delta_{\gamma \delta}+\delta_{\alpha \gamma} \delta_{\beta \delta}+\delta_{\alpha \delta} \delta_{\beta \gamma}\right).
\end{align} 
\end{subequations}
Where $S_d$ is the surface area of the $d$-dimensional unit sphere. For example, $S_2=2 \pi$ and $S_3=4 \pi$. Now to find $A_0$, we integrate \eqref{eq: psi expansion} over the solid angle $\Omega_d$ :
\begin{align}
    &\underbrace{\int \psi d \Omega_d}_c=A_0 c \underbrace{\int d \Omega_d}_{S_d}+A_1 c p_\alpha \underbrace{\int \hat{e}_\alpha d \Omega_d}_0\Rightarrow A_0=\frac{1}{S_d}.
\end{align}
To find $A_1$, we multiply \eqref{eq: psi expansion} by $\hat{e}_\alpha$, then integrate over the solid angle $\Omega_d$:
\begin{align}
    &\underbrace{\int \psi \hat{e}_\alpha d \Omega_d}_{c p_\alpha}=\frac{1}{S_d} c \underbrace{\int \hat{e}_\alpha d \Omega_d}_0+A_1 c p_\beta \underbrace{\int \hat{e}_\alpha \hat{e}_\beta d \Omega_d}_{\frac{S_d}{d} \delta_{\alpha \beta}} \Rightarrow A_1=\frac{d}{S_d}.
\end{align}
Therefore the truncated spherical expansion for $\psi$ can be written as:
\begin{align}
    \psi(\bm{r}, \hat{\bm{e}}, t)=\frac{1}{S_d} c(\bm{r}, t)+\frac{d}{S_d} c(\bm{r}, t) \bm{p}(\bm{r}, t) \cdot \hat{\bm{e}}.\label{eq: psi expression}
\end{align}
We can use \eqref{eq: psi expression} to find the angular averages $\left\langle\hat{e}_\alpha \hat{e}_\beta\right\rangle,\left\langle\hat{e}_\alpha \hat{e}_\beta \hat{e}_\gamma\right\rangle$, and so on. For instance to find $\left\langle\hat{e}_\alpha \hat{e}_\beta\right\rangle$, we use the definition for average:
\begin{align}
    \left\langle\hat{e}_\alpha \hat{e}_\beta\right\rangle&=\int \psi(\bm{r}, \hat{\bm{e}}, t) \hat{e}_\alpha \hat{e}_\beta d \Omega_d.
\end{align}
Which gives:
\begin{subequations}
    \begin{align}
\left\langle\hat{e}_\alpha \hat{e}_\beta\right\rangle&=c\frac{\delta_{\alpha\beta}}{d},\label{eq: average 2}\\
        \left\langle\hat{e}_{\alpha}\hat{e}_{\beta}\hat{e}_{\gamma}\right\rangle&= \frac{c}{d+2}\left(p_{\gamma}\delta_{\alpha\beta} + p_{\beta}\delta_{\alpha\gamma} + p_{\alpha}\delta_{\beta\gamma}\right). \label{eq: average 3}
\end{align}
\end{subequations}

%%%
\section{Non-dimensionalization\label{app:non-dim}}
%%%

In this Appendix, we use the bar notation to indicate non-dimensionalized quantities for clarity, however, in the main text, the reverse is true.
In the following, we consider the dynamics of coupled surfactant-binary
fluid systems in general $d$-dimension. In the simulation, we consider
$d=2$ mostly. 
To get the dimensionless form of the free energy, we introduce $\xi_\text{I}=\sqrt{2\kappa/\beta}$ as the unit of length, and $k_\text{B}T$ as the unit of energy. Dividing Eq.~(\ref{eq:coarse_grained_fe}) by $k_\text{B}T$, and rescaling space $\boldsymbol{r}=\xi_\text{I}\bar{\boldsymbol{r}}$,
the free energy becomes (note that bars indicate dimensionless quantities):
\begin{equation}
\bar{F}=\frac{F}{k_BT}=\int\left\{ -\frac{1}{2}\frac{\beta\xi_{\text{I}}^{d}}{k_\text{B}T}\phi^{2}+\frac{1}{4}\frac{\beta\xi_{\text{I}}^{d}}{k_\text{B}T}\phi^{4}+\frac{1}{2}\frac{\kappa\xi_{\text{I}}^{d-2}}{k_\text{B}T}|\bar{\boldsymbol{\nabla}}\phi|^{2}+\bar{c}\ln(\bar{c})+\frac{d}{2}\bar{c}|\boldsymbol{p}|^{2}+\frac{\chi\ell}{\xi_{\text{I}} k_\text{B}T}\bar{c}\boldsymbol{p}\cdot\bar{\boldsymbol{\nabla}}\phi\right\} d^{d}\bar{r}
\end{equation}
Now we define the dimensionless parameters to be:
\begin{equation}
\bar{\beta}=\frac{\beta\xi_{\text{I}}^{d}}{k_{\text{B}}T},\quad
\bar{\kappa}=\frac{\kappa\xi_{\text{I}}^{d-2}}{k_{\text{B}}T}=\frac{\bar{\beta}}{2},\quad
\text{and}\quad\varepsilon=\frac{\chi\ell}{\xi_{\text{I}} k_{\text{B}}T},
\end{equation}
so that the free energy becomes:
\begin{equation}
\bar{F}=\int\left\{ -\frac{\bar{\beta}}{2}\phi^{2}+\frac{\bar{\beta}}{4}\phi^{4}+\frac{\bar{\beta}}{4}|\bar{\boldsymbol{\nabla}}\phi|^{2}+\bar{c}\ln(\bar{c})+\frac{d}{2}\bar{c}|\boldsymbol{p}|^{2}+\varepsilon\bar{c}\boldsymbol{p}\cdot\bar{\boldsymbol{\nabla}}\phi\right\} d^{d}\bar{r}.
\end{equation}
Now, we introduce $\tau$ as the unit of time, the equations of motion (\ref{eq:dynamics}) become:
\begin{subequations}
\begin{align}
 \frac{\partial \phi}{\partial\bar{t}}+(\bar{\bm{v}}\cdot\bar{\bm{\nabla}}) \phi &= \bar{M} \bar{\bm{\nabla}}^2
    \left[-\bar{\beta}\phi + \bar{\beta}\phi^3 - \frac{\bar{\beta}}{2}\bar{\nabla}^2\phi 
    - \varepsilon\bar{\bm{\nabla}}\cdot(\bar{c}\bm{p}) 
    \right], \\
\frac{\partial\bar{c}}{\partial\bar{t}}+(\bar{\bm{v}}\cdot\bar{\bm{\nabla}})\bar{c} &= \frac{1}{\bar{\gamma}_t}\bar{\nabla}^2\bar{c} + \frac{1}{\bar{\gamma}_t}
\bar{\bm{\nabla}} \cdot 
\left[\bar{c}\bar{\bm{\nabla}}
\left(\frac{d}{2}|\bm{p}|^2 + \varepsilon\bm{p}\cdot\bar{\bm{\nabla}} \phi\right) \right], \\
\frac{\partial\bm{p}}{\partial\bar{t}}+(\bar{\bm{v}}\cdot \bar{\bm{\nabla}})\bm{p} &=  
    -\bar{\bm{\Omega}}\cdot\bm{p} - \frac{Bd}{d+2}\bar{\bm{D}}\cdot\bm{p} 
    - \frac{d-1}{d\bar{\gamma}_r}(d\bm{p} + \varepsilon\bar{\bm{\nabla}}\phi),
\end{align}    
\end{subequations}
where we have introduced the dimensionless parameters: 
\begin{equation}
\bar{M}=\frac{Mk_BT\tau}{\xi_{\text{I}}^{d+2}},\quad
\frac{1}{\bar{\gamma}_{t}}=\frac{k_BT\tau}{\gamma_{t}\xi_{\text{I}}^2},\quad\text{and}\quad
\frac{1}{\bar{\gamma}_{r}}=\frac{k_BT\tau}{\gamma_{r}}.
\end{equation} 
The Stokes equation (\ref{eq:NS_final}) becomes:
\begin{equation}
0 = -\bar{\bm{\nabla}}\bar{P} + \bar{\eta}\bar{\nabla}^2\bar{\bm{v}} + \bar{\bm{f}} + \bar{\bm{\nabla}}\cdot\bar{\bm{\sigma}},
\end{equation}
where $\bar{P}=\xi_{\text{I}}^dP/(k_\text{B}T)$ is the dimensionless pressure and $\bar{\eta}=\eta\xi_{\text{I}}^d/(\tau k_\text{B}T)$ is the dimensionless viscosity.
$\bar{\bm{f}}$ and $\bar{\bm{\sigma}}$ are the dimensionless force density and stress respectively:
\begin{subequations}
\begin{align}
\bar{\bm{f}} &= \frac{\xi^{d+1}_\text{I}}{k_\text{B}T}\bm{f}
= -\phi\bar{\bm{\nabla}}\frac{\delta{\bar{F}}}{\delta\phi}
-\bar{c}\bar{\bm{\nabla}}\frac{\delta{\bar{F}}}{\delta\bar{c}}
-\bm{p}\cdot\left(\bar{\bm{\nabla}}\frac{\delta{\bar{F}}}{\delta\bm{p}}\right)^T \\
\bar{\bm{\sigma}} &= \frac{\xi^d_\text{I}}{k_\text{B}T}\bm{\sigma}
= \frac{Bd}{2(d+2)}\left(\boldsymbol{p}\frac{\delta\bar{F}}{\delta\boldsymbol{p}}+\frac{\delta\bar{F}}{\delta\boldsymbol{p}}\boldsymbol{p}\right)-\frac{1}{2}\left(\boldsymbol{p}\frac{\delta\bar{F}}{\delta\boldsymbol{p}}-\frac{\delta\bar{F}}{\delta\boldsymbol{p}}\boldsymbol{p}\right).
\end{align}    
\end{subequations}

%%%
\section{Numerical boundary conditions\label{app:BC}}
%%%

In our two-dimensional simulation, the space $\bm{r}=(x,y)$ is discretized into $x=i\Delta x$ and $y=j\Delta y$, where $i=0,1,2,\dots N_x-1$ and $j=0,1,2,\dots N_y-1$.
$N_x$ and $N_y$ are the number of lattice points along $x$- and $y$-axis and $\Delta x$ and $\Delta y$ are the corresponding lattice spacing.
The domain size is then $L_x=N_x\Delta x$ along $x$ and $L_y=N_y\Delta y$ along $y$.
We also discretize the time $t$ into $t=n\Delta t$, where $n=0,1,2,\dots$.
The hydrodynamic variables such as $\phi(\bm{r},t),c(\bm{r},t),$ etc. are then discretized into $\phi(\bm{r},t)\rightarrow\phi^n_{ij}$, and so on.
The $\phi$-dynamics, Eq.~\eqref{eq:phidot-nondim} can be discretized into:
\begin{equation}
\phi_{ij}^{n+1} = \phi^n_{ij} - \Delta t\left(\bm{\nabla}\cdot\bm{J}\right)^n_{ij} + \Delta t M \left(\nabla^2\mu\right)^n_{ij},
\end{equation}
where $\bm{J}=\phi\bm{v}$ and $\mu=\delta F/\delta\phi=-\beta\phi+\beta\phi^3-\beta\nabla^2\phi/2-\varepsilon\bm{\nabla}\cdot(c\bm{p})$.
Since $\int\phi(\bm{r},t)\,d^2r$ is conserved, we need to ensure that the discretized $\phi_{ij}^n$ is also conserved, i.e.,
\begin{equation}
\sum_{i=0}^{N_x-1} \sum_{j=0}^{N_y-1} \phi^n_{ij} = \text{constant for all }n\in\mathbb{Z}.
\end{equation}
To ensure this, we discretized $\bm{\nabla}\cdot\bm{J}$ and $\nabla^2\mu$ as follow:
\begin{subequations}
    \begin{align}
        \left(\bm{\nabla}\cdot\bm{J}\right)^n_{ij} &= \frac{J^n_{x,i+1,j}-J^n_{x,i-1,j}}{2\Delta x} + \frac{J^n_{y,i,j+1}-J^n_{y,i,j-1}}{2\Delta y} \\
        \left(\nabla^2\mu\right)^n_{ij} &= \frac{\mu^n_{i+1,j}+\mu^n_{i,j}+\mu^n_{i-1,j}}{\Delta x^2} + \frac{\mu^n_{i,j+1}+\mu^n_{i,j}+\mu^n_{i,j-1}}{\Delta y^2}
    \end{align}
\end{subequations}
We impose the periodic boundary conditions at $j=0$ and $j=N_y-1$ by having:
\begin{equation}
    \bm{J}^n_{i,0} = \bm{J}^n_{i,N_y-1}, \quad \phi^n_{i,0} = \phi^n_{i,N_y-1}, \quad\text{and}\quad \mu^n_{i,0} = \mu^n_{i,N_y-1},
\end{equation}
for all $i$ and $n$.
We impose the no-flux boundary conditions at $i=0$ and $i=N_x-1$ by having:
\begin{align}
    J^n_{x,0,j} &= -J^n_{x,1,j}, \quad J^n_{x,N_x-1,j} = -J^n_{x,N_x-2,j}, \\
    \phi^n_{0,j} &= \phi^n_{1,j}, \quad \phi^n_{N_x-1,j} = \phi^n_{N_x-2,j}, \\
    \mu^n_{0,j} &= \mu^n_{1,j}, \quad\text{and}\quad \mu^n_{N_x-1,j} = \mu^n_{N_x-2,j},
\end{align}
for all $j$ and $n$.
Similar boundary conditions are also used for $c(\bm{r},t)$ and $\bm{p}(\bm{r},t)$, except the no-flux condition does not apply to $\bm{p}$.

\nocite{*}

\bibliography{bibli}% Produces the bibliography via BibTeX.

\begin{thebibliography}{58}%
\makeatletter
\providecommand \@ifxundefined [1]{%
 \@ifx{#1\undefined}
}%
\providecommand \@ifnum [1]{%
 \ifnum #1\expandafter \@firstoftwo
 \else \expandafter \@secondoftwo
 \fi
}%
\providecommand \@ifx [1]{%
 \ifx #1\expandafter \@firstoftwo
 \else \expandafter \@secondoftwo
 \fi
}%
\providecommand \natexlab [1]{#1}%
\providecommand \enquote  [1]{``#1''}%
\providecommand \bibnamefont  [1]{#1}%
\providecommand \bibfnamefont [1]{#1}%
\providecommand \citenamefont [1]{#1}%
\providecommand \href@noop [0]{\@secondoftwo}%
\providecommand \href [0]{\begingroup \@sanitize@url \@href}%
\providecommand \@href[1]{\@@startlink{#1}\@@href}%
\providecommand \@@href[1]{\endgroup#1\@@endlink}%
\providecommand \@sanitize@url [0]{\catcode `\\12\catcode `\$12\catcode
  `\&12\catcode `\#12\catcode `\^12\catcode `\_12\catcode `\%12\relax}%
\providecommand \@@startlink[1]{}%
\providecommand \@@endlink[0]{}%
\providecommand \url  [0]{\begingroup\@sanitize@url \@url }%
\providecommand \@url [1]{\endgroup\@href {#1}{\urlprefix }}%
\providecommand \urlprefix  [0]{URL }%
\providecommand \Eprint [0]{\href }%
\providecommand \doibase [0]{https://doi.org/}%
\providecommand \selectlanguage [0]{\@gobble}%
\providecommand \bibinfo  [0]{\@secondoftwo}%
\providecommand \bibfield  [0]{\@secondoftwo}%
\providecommand \translation [1]{[#1]}%
\providecommand \BibitemOpen [0]{}%
\providecommand \bibitemStop [0]{}%
\providecommand \bibitemNoStop [0]{.\EOS\space}%
\providecommand \EOS [0]{\spacefactor3000\relax}%
\providecommand \BibitemShut  [1]{\csname bibitem#1\endcsname}%
\let\auto@bib@innerbib\@empty
%</preamble>
\bibitem [{\citenamefont {Tadros}(2005)}]{tadros2006applied}%
  \BibitemOpen
  \bibfield  {author} {\bibinfo {author} {\bibfnamefont {T.~F.}\ \bibnamefont
  {Tadros}},\ }\href {https://doi.org/https://doi.org/10.1002/3527604812.ch12}
  {\emph {\bibinfo {title} {Applied Surfactants}}}\ (\bibinfo  {publisher}
  {John Wiley and Sons, Ltd},\ \bibinfo {year} {2005})\ pp.\ \bibinfo {pages}
  {399--432}\BibitemShut {NoStop}%
\bibitem [{\citenamefont {Shaban}\ \emph {et~al.}(2020)\citenamefont {Shaban},
  \citenamefont {Kang},\ and\ \citenamefont {Kim}}]{shaban2020surfactants}%
  \BibitemOpen
  \bibfield  {author} {\bibinfo {author} {\bibfnamefont {S.~M.}\ \bibnamefont
  {Shaban}}, \bibinfo {author} {\bibfnamefont {J.}~\bibnamefont {Kang}},\ and\
  \bibinfo {author} {\bibfnamefont {D.-H.}\ \bibnamefont {Kim}},\ }\bibfield
  {title} {\bibinfo {title} {Surfactants: Recent advances and their
  applications},\ }\href
  {https://doi.org/https://doi.org/10.1016/j.coco.2020.100537} {\bibfield
  {journal} {\bibinfo  {journal} {Composites Communications}\ }\textbf
  {\bibinfo {volume} {22}},\ \bibinfo {pages} {100537} (\bibinfo {year}
  {2020})}\BibitemShut {NoStop}%
\bibitem [{\citenamefont {Mulqueen}\ and\ \citenamefont
  {Blankschtein}(2002)}]{mulqueen2002theoretical}%
  \BibitemOpen
  \bibfield  {author} {\bibinfo {author} {\bibfnamefont {M.}~\bibnamefont
  {Mulqueen}}\ and\ \bibinfo {author} {\bibfnamefont {D.}~\bibnamefont
  {Blankschtein}},\ }\bibfield  {title} {\bibinfo {title} {Theoretical and
  experimental investigation of the equilibrium oil-water interfacial tensions
  of solutions containing surfactant mixtures},\ }\href
  {https://doi.org/10.1021/la010993u} {\bibfield  {journal} {\bibinfo
  {journal} {Langmuir}\ }\textbf {\bibinfo {volume} {18}},\ \bibinfo {pages}
  {365} (\bibinfo {year} {2002})}\BibitemShut {NoStop}%
\bibitem [{\citenamefont {Wang}\ \emph {et~al.}(2017)\citenamefont {Wang},
  \citenamefont {Haghmoradi}, \citenamefont {Liu}, \citenamefont {Xi},
  \citenamefont {Hirasaki}, \citenamefont {Miller},\ and\ \citenamefont
  {Chapman}}]{wang2017modeling}%
  \BibitemOpen
  \bibfield  {author} {\bibinfo {author} {\bibfnamefont {L.}~\bibnamefont
  {Wang}}, \bibinfo {author} {\bibfnamefont {A.}~\bibnamefont {Haghmoradi}},
  \bibinfo {author} {\bibfnamefont {J.}~\bibnamefont {Liu}}, \bibinfo {author}
  {\bibfnamefont {S.}~\bibnamefont {Xi}}, \bibinfo {author} {\bibfnamefont
  {G.~J.}\ \bibnamefont {Hirasaki}}, \bibinfo {author} {\bibfnamefont {C.~A.}\
  \bibnamefont {Miller}},\ and\ \bibinfo {author} {\bibfnamefont {W.~G.}\
  \bibnamefont {Chapman}},\ }\bibfield  {title} {\bibinfo {title} {Modeling
  micelle formation and interfacial properties with isaft classical density
  functional theory},\ }\href {https://doi.org/10.1063/1.4978503} {\bibfield
  {journal} {\bibinfo  {journal} {The Journal of Chemical Physics}\ }\textbf
  {\bibinfo {volume} {146}},\ \bibinfo {pages} {124705} (\bibinfo {year}
  {2017})}\BibitemShut {NoStop}%
\bibitem [{\citenamefont {Touhami}\ \emph {et~al.}(2001)\citenamefont
  {Touhami}, \citenamefont {Rana}, \citenamefont {Neale},\ and\ \citenamefont
  {Hornof}}]{touhami2001study}%
  \BibitemOpen
  \bibfield  {author} {\bibinfo {author} {\bibfnamefont {Y.}~\bibnamefont
  {Touhami}}, \bibinfo {author} {\bibfnamefont {D.}~\bibnamefont {Rana}},
  \bibinfo {author} {\bibfnamefont {G.~H.}\ \bibnamefont {Neale}},\ and\
  \bibinfo {author} {\bibfnamefont {V.}~\bibnamefont {Hornof}},\ }\bibfield
  {title} {\bibinfo {title} {Study of polymer-surfactant interactions via
  surface tension measurements},\ }\href
  {https://doi.org/10.1007/s003960000455} {\bibfield  {journal} {\bibinfo
  {journal} {Colloid and Polymer Science}\ }\textbf {\bibinfo {volume} {279}},\
  \bibinfo {pages} {297} (\bibinfo {year} {2001})}\BibitemShut {NoStop}%
\bibitem [{\citenamefont {Santos}\ and\ \citenamefont
  {Panagiotopoulos}(2016)}]{santos2016determination}%
  \BibitemOpen
  \bibfield  {author} {\bibinfo {author} {\bibfnamefont {A.~P.}\ \bibnamefont
  {Santos}}\ and\ \bibinfo {author} {\bibfnamefont {A.~Z.}\ \bibnamefont
  {Panagiotopoulos}},\ }\bibfield  {title} {\bibinfo {title} {Determination of
  the critical micelle concentration in simulations of surfactant systems},\
  }\href {https://doi.org/10.1063/1.4940687} {\bibfield  {journal} {\bibinfo
  {journal} {The Journal of Chemical Physics}\ }\textbf {\bibinfo {volume}
  {144}},\ \bibinfo {pages} {044709} (\bibinfo {year} {2016})}\BibitemShut
  {NoStop}%
\bibitem [{\citenamefont {Gangula}\ \emph {et~al.}(2010)\citenamefont
  {Gangula}, \citenamefont {Suen},\ and\ \citenamefont
  {Conte}}]{gangula2010analytical}%
  \BibitemOpen
  \bibfield  {author} {\bibinfo {author} {\bibfnamefont {S.}~\bibnamefont
  {Gangula}}, \bibinfo {author} {\bibfnamefont {S.-Y.}\ \bibnamefont {Suen}},\
  and\ \bibinfo {author} {\bibfnamefont {E.~D.}\ \bibnamefont {Conte}},\
  }\bibfield  {title} {\bibinfo {title} {Analytical applications of admicelle
  and hemimicelle solid phase extraction of organic analytes},\ }\href
  {https://doi.org/https://doi.org/10.1016/j.microc.2009.10.005} {\bibfield
  {journal} {\bibinfo  {journal} {Microchemical Journal}\ }\textbf {\bibinfo
  {volume} {95}},\ \bibinfo {pages} {2} (\bibinfo {year} {2010})}\BibitemShut
  {NoStop}%
\bibitem [{\citenamefont {Krebs}\ \emph {et~al.}(2012)\citenamefont {Krebs},
  \citenamefont {Schroën},\ and\ \citenamefont {Boom}}]{krebs2012coalescence}%
  \BibitemOpen
  \bibfield  {author} {\bibinfo {author} {\bibfnamefont {T.}~\bibnamefont
  {Krebs}}, \bibinfo {author} {\bibfnamefont {K.}~\bibnamefont {Schroën}},\
  and\ \bibinfo {author} {\bibfnamefont {R.}~\bibnamefont {Boom}},\ }\bibfield
  {title} {\bibinfo {title} {Coalescence dynamics of surfactant-stabilized
  emulsions studied with microfluidics},\ }\href
  {https://doi.org/10.1039/C2SM26122G} {\bibfield  {journal} {\bibinfo
  {journal} {Soft Matter}\ }\textbf {\bibinfo {volume} {8}},\ \bibinfo {pages}
  {10650} (\bibinfo {year} {2012})}\BibitemShut {NoStop}%
\bibitem [{\citenamefont {Dai}\ and\ \citenamefont
  {Leal}(2008)}]{dai2008mechanism}%
  \BibitemOpen
  \bibfield  {author} {\bibinfo {author} {\bibfnamefont {B.}~\bibnamefont
  {Dai}}\ and\ \bibinfo {author} {\bibfnamefont {L.~G.}\ \bibnamefont {Leal}},\
  }\bibfield  {title} {\bibinfo {title} {The mechanism of surfactant effects on
  drop coalescence},\ }\href {https://doi.org/10.1063/1.2911700} {\bibfield
  {journal} {\bibinfo  {journal} {Physics of Fluids}\ }\textbf {\bibinfo
  {volume} {20}},\ \bibinfo {pages} {040802} (\bibinfo {year}
  {2008})}\BibitemShut {NoStop}%
\bibitem [{\citenamefont {{Alexander, S.}}(1978)}]{alexander1978lattice}%
  \BibitemOpen
  \bibfield  {author} {\bibinfo {author} {\bibnamefont {{Alexander, S.}}},\
  }\bibfield  {title} {\bibinfo {title} {A lattice gas model for
  microemulsions},\ }\href {https://doi.org/10.1051/jphyslet:019780039010100}
  {\bibfield  {journal} {\bibinfo  {journal} {J. Physique Lett.}\ }\textbf
  {\bibinfo {volume} {39}},\ \bibinfo {pages} {1} (\bibinfo {year}
  {1978})}\BibitemShut {NoStop}%
\bibitem [{\citenamefont {Ahluwalia}\ and\ \citenamefont
  {Puri}(1996)}]{ahluwalia1996}%
  \BibitemOpen
  \bibfield  {author} {\bibinfo {author} {\bibfnamefont {R.}~\bibnamefont
  {Ahluwalia}}\ and\ \bibinfo {author} {\bibfnamefont {S.}~\bibnamefont
  {Puri}},\ }\bibfield  {title} {\bibinfo {title} {Phase-ordering dynamics in
  binary mixtures with surfactants},\ }\href
  {https://doi.org/10.1088/0953-8984/8/3/004} {\bibfield  {journal} {\bibinfo
  {journal} {Journal of Physics: Condensed Matter}\ }\textbf {\bibinfo {volume}
  {8}},\ \bibinfo {pages} {227} (\bibinfo {year} {1996})}\BibitemShut {NoStop}%
\bibitem [{\citenamefont {Gilhøj}\ \emph {et~al.}(1996)\citenamefont
  {Gilhøj}, \citenamefont {Laradji}, \citenamefont {Dammann}, \citenamefont
  {Jeppesen}, \citenamefont {Mouritsen}, \citenamefont {Toxværd},\ and\
  \citenamefont {Zuckermann}}]{gilhoj1996}%
  \BibitemOpen
  \bibfield  {author} {\bibinfo {author} {\bibfnamefont {H.}~\bibnamefont
  {Gilhøj}}, \bibinfo {author} {\bibfnamefont {M.}~\bibnamefont {Laradji}},
  \bibinfo {author} {\bibfnamefont {B.}~\bibnamefont {Dammann}}, \bibinfo
  {author} {\bibfnamefont {C.}~\bibnamefont {Jeppesen}}, \bibinfo {author}
  {\bibfnamefont {O.~G.}\ \bibnamefont {Mouritsen}}, \bibinfo {author}
  {\bibfnamefont {S.}~\bibnamefont {Toxværd}},\ and\ \bibinfo {author}
  {\bibfnamefont {M.~J.}\ \bibnamefont {Zuckermann}},\ }\bibfield  {title}
  {\bibinfo {title} {Effect of vacancies and surfactants on the dynamics of
  ordering processes in multi-component systems},\ }\href
  {https://doi.org/https://doi.org/10.1016/0378-4754(95)00041-0} {\bibfield
  {journal} {\bibinfo  {journal} {Mathematics and Computers in Simulation}\
  }\textbf {\bibinfo {volume} {40}},\ \bibinfo {pages} {319} (\bibinfo {year}
  {1996})}\BibitemShut {NoStop}%
\bibitem [{\citenamefont {Laradji}\ and\ \citenamefont
  {Mouritsen}(2000)}]{laradji2000md}%
  \BibitemOpen
  \bibfield  {author} {\bibinfo {author} {\bibfnamefont {M.}~\bibnamefont
  {Laradji}}\ and\ \bibinfo {author} {\bibfnamefont {O.~G.}\ \bibnamefont
  {Mouritsen}},\ }\bibfield  {title} {\bibinfo {title} {Elastic properties of
  surfactant monolayers at liquid–liquid interfaces: A molecular dynamics
  study},\ }\href {https://doi.org/10.1063/1.481486} {\bibfield  {journal}
  {\bibinfo  {journal} {The Journal of Chemical Physics}\ }\textbf {\bibinfo
  {volume} {112}},\ \bibinfo {pages} {8621} (\bibinfo {year}
  {2000})}\BibitemShut {NoStop}%
\bibitem [{\citenamefont {Kawasaki}\ and\ \citenamefont
  {Kawakatsu}(1990)}]{kawasaki_continuum_1990}%
  \BibitemOpen
  \bibfield  {author} {\bibinfo {author} {\bibfnamefont {K.}~\bibnamefont
  {Kawasaki}}\ and\ \bibinfo {author} {\bibfnamefont {T.}~\bibnamefont
  {Kawakatsu}},\ }\bibfield  {title} {{\selectlanguage {en}\bibinfo {title}
  {Continuum theory of an immiscible binary fluid mixture with a surfactant}},\
  }\href {https://doi.org/10.1016/0378-4371(90)90222-E} {\bibfield  {journal}
  {\bibinfo  {journal} {Physica A: Statistical Mechanics and its Applications}\
  }\textbf {\bibinfo {volume} {164}},\ \bibinfo {pages} {549} (\bibinfo {year}
  {1990})}\BibitemShut {NoStop}%
\bibitem [{\citenamefont {Anisimov}\ \emph {et~al.}(1992)\citenamefont
  {Anisimov}, \citenamefont {Gorodetsky}, \citenamefont {Davydov},\ and\
  \citenamefont {Kurliandsky}}]{anisimov_landau_1992}%
  \BibitemOpen
  \bibfield  {author} {\bibinfo {author} {\bibfnamefont {M.~A.}\ \bibnamefont
  {Anisimov}}, \bibinfo {author} {\bibfnamefont {E.~E.}\ \bibnamefont
  {Gorodetsky}}, \bibinfo {author} {\bibfnamefont {A.~J.}\ \bibnamefont
  {Davydov}},\ and\ \bibinfo {author} {\bibfnamefont {A.~S.}\ \bibnamefont
  {Kurliandsky}},\ }\bibfield  {title} {{\selectlanguage {en}\bibinfo {title}
  {Landau model for self-assembly and liquid crystal formation in surfactant
  solutions}},\ }\href {https://doi.org/10.1080/02678299208030697} {\bibfield
  {journal} {\bibinfo  {journal} {Liquid Crystals}\ }\textbf {\bibinfo {volume}
  {11}},\ \bibinfo {pages} {941} (\bibinfo {year} {1992})}\BibitemShut
  {NoStop}%
\bibitem [{\citenamefont {Melenkevitz}\ and\ \citenamefont
  {Javadpour}(1997)}]{melenkevitz_phase_1997}%
  \BibitemOpen
  \bibfield  {author} {\bibinfo {author} {\bibfnamefont {J.}~\bibnamefont
  {Melenkevitz}}\ and\ \bibinfo {author} {\bibfnamefont {S.~H.}\ \bibnamefont
  {Javadpour}},\ }\bibfield  {title} {{\selectlanguage {en}\bibinfo {title}
  {Phase separation dynamics in mixtures containing surfactants}},\ }\href
  {https://doi.org/10.1063/1.474422} {\bibfield  {journal} {\bibinfo  {journal}
  {The Journal of Chemical Physics}\ }\textbf {\bibinfo {volume} {107}},\
  \bibinfo {pages} {623} (\bibinfo {year} {1997})}\BibitemShut {NoStop}%
\bibitem [{\citenamefont {Zhu}\ \emph {et~al.}(2020)\citenamefont {Zhu},
  \citenamefont {Kou}, \citenamefont {Yao}, \citenamefont {Li},\ and\
  \citenamefont {Sun}}]{zhu2020phase}%
  \BibitemOpen
  \bibfield  {author} {\bibinfo {author} {\bibfnamefont {G.}~\bibnamefont
  {Zhu}}, \bibinfo {author} {\bibfnamefont {J.}~\bibnamefont {Kou}}, \bibinfo
  {author} {\bibfnamefont {J.}~\bibnamefont {Yao}}, \bibinfo {author}
  {\bibfnamefont {A.}~\bibnamefont {Li}},\ and\ \bibinfo {author}
  {\bibfnamefont {S.}~\bibnamefont {Sun}},\ }\bibfield  {title} {\bibinfo
  {title} {A phase-field moving contact line model with soluble surfactants},\
  }\href {https://doi.org/https://doi.org/10.1016/j.jcp.2019.109170} {\bibfield
   {journal} {\bibinfo  {journal} {Journal of Computational Physics}\ }\textbf
  {\bibinfo {volume} {405}},\ \bibinfo {pages} {109170} (\bibinfo {year}
  {2020})}\BibitemShut {NoStop}%
\bibitem [{\citenamefont {Liu}\ and\ \citenamefont
  {Zhang}(2010)}]{liu2010phase}%
  \BibitemOpen
  \bibfield  {author} {\bibinfo {author} {\bibfnamefont {H.}~\bibnamefont
  {Liu}}\ and\ \bibinfo {author} {\bibfnamefont {Y.}~\bibnamefont {Zhang}},\
  }\bibfield  {title} {\bibinfo {title} {Phase-field modeling droplet dynamics
  with soluble surfactants},\ }\href
  {https://doi.org/https://doi.org/10.1016/j.jcp.2010.08.031} {\bibfield
  {journal} {\bibinfo  {journal} {Journal of Computational Physics}\ }\textbf
  {\bibinfo {volume} {229}},\ \bibinfo {pages} {9166} (\bibinfo {year}
  {2010})}\BibitemShut {NoStop}%
\bibitem [{\citenamefont {Kalam}\ \emph {et~al.}(2021)\citenamefont {Kalam},
  \citenamefont {Abu-Khamsin}, \citenamefont {Kamal},\ and\ \citenamefont
  {Patil}}]{kalam2021surfactant}%
  \BibitemOpen
  \bibfield  {author} {\bibinfo {author} {\bibfnamefont {S.}~\bibnamefont
  {Kalam}}, \bibinfo {author} {\bibfnamefont {S.~A.}\ \bibnamefont
  {Abu-Khamsin}}, \bibinfo {author} {\bibfnamefont {M.~S.}\ \bibnamefont
  {Kamal}},\ and\ \bibinfo {author} {\bibfnamefont {S.}~\bibnamefont {Patil}},\
  }\bibfield  {title} {\bibinfo {title} {Surfactant adsorption isotherms: A
  review},\ }\href {https://doi.org/10.1021/acsomega.1c04661} {\bibfield
  {journal} {\bibinfo  {journal} {ACS Omega}\ }\textbf {\bibinfo {volume}
  {6}},\ \bibinfo {pages} {32342} (\bibinfo {year} {2021})}\BibitemShut
  {NoStop}%
\bibitem [{\citenamefont {Manikantan}\ and\ \citenamefont
  {Squires}(2020)}]{manikantan2020surfactant}%
  \BibitemOpen
  \bibfield  {author} {\bibinfo {author} {\bibfnamefont {H.}~\bibnamefont
  {Manikantan}}\ and\ \bibinfo {author} {\bibfnamefont {T.~M.}\ \bibnamefont
  {Squires}},\ }\bibfield  {title} {\bibinfo {title} {Surfactant dynamics:
  hidden variables controlling fluid flows},\ }\href
  {https://doi.org/10.1017/jfm.2020.170} {\bibfield  {journal} {\bibinfo
  {journal} {Journal of Fluid Mechanics}\ }\textbf {\bibinfo {volume} {892}},\
  \bibinfo {pages} {P1} (\bibinfo {year} {2020})}\BibitemShut {NoStop}%
\bibitem [{\citenamefont {Soligo}\ \emph
  {et~al.}(2019{\natexlab{a}})\citenamefont {Soligo}, \citenamefont {Roccon},\
  and\ \citenamefont {Soldati}}]{soligo2019coalescence}%
  \BibitemOpen
  \bibfield  {author} {\bibinfo {author} {\bibfnamefont {G.}~\bibnamefont
  {Soligo}}, \bibinfo {author} {\bibfnamefont {A.}~\bibnamefont {Roccon}},\
  and\ \bibinfo {author} {\bibfnamefont {A.}~\bibnamefont {Soldati}},\
  }\bibfield  {title} {\bibinfo {title} {Coalescence of surfactant-laden drops
  by phase field method},\ }\href
  {https://doi.org/https://doi.org/10.1016/j.jcp.2018.10.021} {\bibfield
  {journal} {\bibinfo  {journal} {Journal of Computational Physics}\ }\textbf
  {\bibinfo {volume} {376}},\ \bibinfo {pages} {1292} (\bibinfo {year}
  {2019}{\natexlab{a}})}\BibitemShut {NoStop}%
\bibitem [{\citenamefont {Soligo}\ \emph
  {et~al.}(2019{\natexlab{b}})\citenamefont {Soligo}, \citenamefont {Roccon},\
  and\ \citenamefont {Soldati}}]{soligo2019breakage}%
  \BibitemOpen
  \bibfield  {author} {\bibinfo {author} {\bibfnamefont {G.}~\bibnamefont
  {Soligo}}, \bibinfo {author} {\bibfnamefont {A.}~\bibnamefont {Roccon}},\
  and\ \bibinfo {author} {\bibfnamefont {A.}~\bibnamefont {Soldati}},\
  }\bibfield  {title} {\bibinfo {title} {Breakage, coalescence and size
  distribution of surfactant-laden droplets in turbulent flow},\ }\href
  {https://doi.org/10.1017/jfm.2019.772} {\bibfield  {journal} {\bibinfo
  {journal} {Journal of Fluid Mechanics}\ }\textbf {\bibinfo {volume} {881}},\
  \bibinfo {pages} {244} (\bibinfo {year} {2019}{\natexlab{b}})}\BibitemShut
  {NoStop}%
\bibitem [{\citenamefont {Yamashita}\ \emph {et~al.}(2024)\citenamefont
  {Yamashita}, \citenamefont {Matsushita},\ and\ \citenamefont
  {Suekane}}]{yamashita2024conservative}%
  \BibitemOpen
  \bibfield  {author} {\bibinfo {author} {\bibfnamefont {S.}~\bibnamefont
  {Yamashita}}, \bibinfo {author} {\bibfnamefont {S.}~\bibnamefont
  {Matsushita}},\ and\ \bibinfo {author} {\bibfnamefont {T.}~\bibnamefont
  {Suekane}},\ }\bibfield  {title} {\bibinfo {title} {Conservative transport
  model for surfactant on the interface based on the phase-field method},\
  }\href {https://doi.org/https://doi.org/10.1016/j.jcp.2024.113292} {\bibfield
   {journal} {\bibinfo  {journal} {Journal of Computational Physics}\ }\textbf
  {\bibinfo {volume} {516}},\ \bibinfo {pages} {113292} (\bibinfo {year}
  {2024})}\BibitemShut {NoStop}%
\bibitem [{\citenamefont {{Erik Teigen}}\ \emph {et~al.}(2011)\citenamefont
  {{Erik Teigen}}, \citenamefont {Song}, \citenamefont {Lowengrub},\ and\
  \citenamefont {Voigt}}]{teigen2011diffuse}%
  \BibitemOpen
  \bibfield  {author} {\bibinfo {author} {\bibfnamefont {K.}~\bibnamefont
  {{Erik Teigen}}}, \bibinfo {author} {\bibfnamefont {P.}~\bibnamefont {Song}},
  \bibinfo {author} {\bibfnamefont {J.}~\bibnamefont {Lowengrub}},\ and\
  \bibinfo {author} {\bibfnamefont {A.}~\bibnamefont {Voigt}},\ }\bibfield
  {title} {\bibinfo {title} {A diffuse-interface method for two-phase flows
  with soluble surfactants},\ }\href
  {https://doi.org/https://doi.org/10.1016/j.jcp.2010.09.020} {\bibfield
  {journal} {\bibinfo  {journal} {Journal of Computational Physics}\ }\textbf
  {\bibinfo {volume} {230}},\ \bibinfo {pages} {375} (\bibinfo {year}
  {2011})}\BibitemShut {NoStop}%
\bibitem [{\citenamefont {Yang}(2021)}]{yang2021novel}%
  \BibitemOpen
  \bibfield  {author} {\bibinfo {author} {\bibfnamefont {X.}~\bibnamefont
  {Yang}},\ }\bibfield  {title} {\bibinfo {title} {A novel fully-decoupled,
  second-order and energy stable numerical scheme of the conserved allen–cahn
  type flow-coupled binary surfactant model},\ }\href
  {https://doi.org/https://doi.org/10.1016/j.cma.2020.113502} {\bibfield
  {journal} {\bibinfo  {journal} {Computer Methods in Applied Mechanics and
  Engineering}\ }\textbf {\bibinfo {volume} {373}},\ \bibinfo {pages} {113502}
  (\bibinfo {year} {2021})}\BibitemShut {NoStop}%
\bibitem [{\citenamefont {Booty}\ and\ \citenamefont
  {Siegel}(2010)}]{booty2010hybrid}%
  \BibitemOpen
  \bibfield  {author} {\bibinfo {author} {\bibfnamefont {M.}~\bibnamefont
  {Booty}}\ and\ \bibinfo {author} {\bibfnamefont {M.}~\bibnamefont {Siegel}},\
  }\bibfield  {title} {\bibinfo {title} {A hybrid numerical method for
  interfacial fluid flow with soluble surfactant},\ }\href
  {https://doi.org/https://doi.org/10.1016/j.jcp.2010.01.032} {\bibfield
  {journal} {\bibinfo  {journal} {Journal of Computational Physics}\ }\textbf
  {\bibinfo {volume} {229}},\ \bibinfo {pages} {3864} (\bibinfo {year}
  {2010})}\BibitemShut {NoStop}%
\bibitem [{\citenamefont {Kim}(2006)}]{kim2006numerical}%
  \BibitemOpen
  \bibfield  {author} {\bibinfo {author} {\bibfnamefont {J.}~\bibnamefont
  {Kim}},\ }\bibfield  {title} {\bibinfo {title} {Numerical simulations of
  phase separation dynamics in a water-oil-surfactant system},\ }\href
  {https://doi.org/https://doi.org/10.1016/j.jcis.2006.07.032} {\bibfield
  {journal} {\bibinfo  {journal} {Journal of Colloid and Interface Science}\
  }\textbf {\bibinfo {volume} {303}},\ \bibinfo {pages} {272} (\bibinfo {year}
  {2006})}\BibitemShut {NoStop}%
\bibitem [{\citenamefont {Ganesan}(2015)}]{ganesan2015simulations}%
  \BibitemOpen
  \bibfield  {author} {\bibinfo {author} {\bibfnamefont {S.}~\bibnamefont
  {Ganesan}},\ }\bibfield  {title} {\bibinfo {title} {Simulations of impinging
  droplets with surfactant-dependent dynamic contact angle},\ }\href
  {https://doi.org/https://doi.org/10.1016/j.jcp.2015.08.026} {\bibfield
  {journal} {\bibinfo  {journal} {Journal of Computational Physics}\ }\textbf
  {\bibinfo {volume} {301}},\ \bibinfo {pages} {178} (\bibinfo {year}
  {2015})}\BibitemShut {NoStop}%
\bibitem [{\citenamefont {Frisch}\ \emph {et~al.}(1986)\citenamefont {Frisch},
  \citenamefont {Hasslacher},\ and\ \citenamefont
  {Pomeau}}]{PhysRevLett.56.1505}%
  \BibitemOpen
  \bibfield  {author} {\bibinfo {author} {\bibfnamefont {U.}~\bibnamefont
  {Frisch}}, \bibinfo {author} {\bibfnamefont {B.}~\bibnamefont {Hasslacher}},\
  and\ \bibinfo {author} {\bibfnamefont {Y.}~\bibnamefont {Pomeau}},\
  }\bibfield  {title} {\bibinfo {title} {Lattice-gas automata for the
  navier-stokes equation},\ }\href
  {https://doi.org/10.1103/PhysRevLett.56.1505} {\bibfield  {journal} {\bibinfo
   {journal} {Phys. Rev. Lett.}\ }\textbf {\bibinfo {volume} {56}},\ \bibinfo
  {pages} {1505} (\bibinfo {year} {1986})}\BibitemShut {NoStop}%
\bibitem [{\citenamefont {Higuera}\ \emph {et~al.}(1989)\citenamefont
  {Higuera}, \citenamefont {Succi},\ and\ \citenamefont
  {Benzi}}]{higuera_lattice_1989}%
  \BibitemOpen
  \bibfield  {author} {\bibinfo {author} {\bibfnamefont {F.~J.}\ \bibnamefont
  {Higuera}}, \bibinfo {author} {\bibfnamefont {S.}~\bibnamefont {Succi}},\
  and\ \bibinfo {author} {\bibfnamefont {R.}~\bibnamefont {Benzi}},\ }\bibfield
   {title} {{\selectlanguage {en}\bibinfo {title} {Lattice {Gas} {Dynamics}
  with {Enhanced} {Collisions}}},\ }\href
  {https://doi.org/10.1209/0295-5075/9/4/008} {\bibfield  {journal} {\bibinfo
  {journal} {Europhysics Letters (EPL)}\ }\textbf {\bibinfo {volume} {9}},\
  \bibinfo {pages} {345} (\bibinfo {year} {1989})}\BibitemShut {NoStop}%
\bibitem [{\citenamefont {Benzi}\ \emph {et~al.}(1992)\citenamefont {Benzi},
  \citenamefont {Succi},\ and\ \citenamefont
  {Vergassola}}]{benzi_lattice_1992}%
  \BibitemOpen
  \bibfield  {author} {\bibinfo {author} {\bibfnamefont {R.}~\bibnamefont
  {Benzi}}, \bibinfo {author} {\bibfnamefont {S.}~\bibnamefont {Succi}},\ and\
  \bibinfo {author} {\bibfnamefont {M.}~\bibnamefont {Vergassola}},\ }\bibfield
   {title} {{\selectlanguage {en}\bibinfo {title} {The lattice {Boltzmann}
  equation: theory and applications}},\ }\href
  {https://doi.org/10.1016/0370-1573(92)90090-M} {\bibfield  {journal}
  {\bibinfo  {journal} {Physics Reports}\ }\textbf {\bibinfo {volume} {222}},\
  \bibinfo {pages} {145} (\bibinfo {year} {1992})}\BibitemShut {NoStop}%
\bibitem [{\citenamefont {Rothman}\ and\ \citenamefont
  {Keller}(1988)}]{rothman_immiscible_1988}%
  \BibitemOpen
  \bibfield  {author} {\bibinfo {author} {\bibfnamefont {D.~H.}\ \bibnamefont
  {Rothman}}\ and\ \bibinfo {author} {\bibfnamefont {J.~M.}\ \bibnamefont
  {Keller}},\ }\bibfield  {title} {{\selectlanguage {en}\bibinfo {title}
  {Immiscible cellular-automaton fluids}},\ }\href
  {https://doi.org/10.1007/BF01019743} {\bibfield  {journal} {\bibinfo
  {journal} {Journal of Statistical Physics}\ }\textbf {\bibinfo {volume}
  {52}},\ \bibinfo {pages} {1119} (\bibinfo {year} {1988})}\BibitemShut
  {NoStop}%
\bibitem [{\citenamefont {Chan}\ and\ \citenamefont
  {Liang}(1990)}]{chan_critical_1990}%
  \BibitemOpen
  \bibfield  {author} {\bibinfo {author} {\bibfnamefont {C.~K.}\ \bibnamefont
  {Chan}}\ and\ \bibinfo {author} {\bibfnamefont {N.~Y.}\ \bibnamefont
  {Liang}},\ }\bibfield  {title} {{\selectlanguage {en}\bibinfo {title}
  {Critical {Phenomena} in an {Immiscible} {Lattice}-{Gas} {Cellular}
  {Automaton}}},\ }\href {https://doi.org/10.1209/0295-5075/13/6/004}
  {\bibfield  {journal} {\bibinfo  {journal} {Europhysics Letters (EPL)}\
  }\textbf {\bibinfo {volume} {13}},\ \bibinfo {pages} {495} (\bibinfo {year}
  {1990})}\BibitemShut {NoStop}%
\bibitem [{\citenamefont {Love}\ \emph {et~al.}(2003)\citenamefont {Love},
  \citenamefont {Nekovee}, \citenamefont {Coveney}, \citenamefont {Chin},
  \citenamefont {González-Segredo},\ and\ \citenamefont
  {Martin}}]{love2003multiphase}%
  \BibitemOpen
  \bibfield  {author} {\bibinfo {author} {\bibfnamefont {P.}~\bibnamefont
  {Love}}, \bibinfo {author} {\bibfnamefont {M.}~\bibnamefont {Nekovee}},
  \bibinfo {author} {\bibfnamefont {P.}~\bibnamefont {Coveney}}, \bibinfo
  {author} {\bibfnamefont {J.}~\bibnamefont {Chin}}, \bibinfo {author}
  {\bibfnamefont {N.}~\bibnamefont {González-Segredo}},\ and\ \bibinfo
  {author} {\bibfnamefont {J.}~\bibnamefont {Martin}},\ }\bibfield  {title}
  {\bibinfo {title} {Simulations of amphiphilic fluids using mesoscale
  lattice-boltzmann and lattice-gas methods},\ }\href
  {https://doi.org/https://doi.org/10.1016/S0010-4655(03)00200-5} {\bibfield
  {journal} {\bibinfo  {journal} {Computer Physics Communications}\ }\textbf
  {\bibinfo {volume} {153}},\ \bibinfo {pages} {340} (\bibinfo {year}
  {2003})}\BibitemShut {NoStop}%
\bibitem [{\citenamefont {Theissen}\ and\ \citenamefont
  {Gompper}(1999)}]{theissen1999lattice}%
  \BibitemOpen
  \bibfield  {author} {\bibinfo {author} {\bibfnamefont {O.}~\bibnamefont
  {Theissen}}\ and\ \bibinfo {author} {\bibfnamefont {G.}~\bibnamefont
  {Gompper}},\ }\bibfield  {title} {\bibinfo {title} {Lattice-boltzmann study
  of spontaneous emulsification},\ }\href
  {https://doi.org/10.1007/s100510050920} {\bibfield  {journal} {\bibinfo
  {journal} {The European Physical Journal B - Condensed Matter and Complex
  Systems}\ }\textbf {\bibinfo {volume} {11}},\ \bibinfo {pages} {91} (\bibinfo
  {year} {1999})}\BibitemShut {NoStop}%
\bibitem [{\citenamefont {Kian~Far}\ \emph {et~al.}(2021)\citenamefont
  {Kian~Far}, \citenamefont {Gorakifard},\ and\ \citenamefont
  {Fattahi}}]{kian2021multiphase}%
  \BibitemOpen
  \bibfield  {author} {\bibinfo {author} {\bibfnamefont {E.}~\bibnamefont
  {Kian~Far}}, \bibinfo {author} {\bibfnamefont {M.}~\bibnamefont
  {Gorakifard}},\ and\ \bibinfo {author} {\bibfnamefont {E.}~\bibnamefont
  {Fattahi}},\ }\bibfield  {title} {\bibinfo {title} {Multiphase phase-field
  lattice boltzmann method for simulation of soluble surfactants},\ }\href
  {https://doi.org/10.3390/sym13061019} {\bibfield  {journal} {\bibinfo
  {journal} {Symmetry}\ }\textbf {\bibinfo {volume} {13}},\ \bibinfo {pages}
  {1019} (\bibinfo {year} {2021})}\BibitemShut {NoStop}%
\bibitem [{\citenamefont {Krüger}\ \emph {et~al.}(2017)\citenamefont
  {Krüger}, \citenamefont {Kusumaatmaja}, \citenamefont {Kuzmin},
  \citenamefont {Shardt}, \citenamefont {Silva},\ and\ \citenamefont
  {Viggen}}]{kruger_lattice_2017}%
  \BibitemOpen
  \bibfield  {author} {\bibinfo {author} {\bibfnamefont {T.}~\bibnamefont
  {Krüger}}, \bibinfo {author} {\bibfnamefont {H.}~\bibnamefont
  {Kusumaatmaja}}, \bibinfo {author} {\bibfnamefont {A.}~\bibnamefont
  {Kuzmin}}, \bibinfo {author} {\bibfnamefont {O.}~\bibnamefont {Shardt}},
  \bibinfo {author} {\bibfnamefont {G.}~\bibnamefont {Silva}},\ and\ \bibinfo
  {author} {\bibfnamefont {E.~M.}\ \bibnamefont {Viggen}},\ }\href
  {https://doi.org/10.1007/978-3-319-44649-3} {{\selectlanguage {en}\emph
  {\bibinfo {title} {The {Lattice} {Boltzmann} {Method}: {Principles} and
  {Practice}}}}},\ Graduate {Texts} in {Physics}\ (\bibinfo  {publisher}
  {Springer International Publishing},\ \bibinfo {address} {Cham},\ \bibinfo
  {year} {2017})\BibitemShut {NoStop}%
\bibitem [{\citenamefont {Pelusi}\ \emph {et~al.}(2022)\citenamefont {Pelusi},
  \citenamefont {Lulli}, \citenamefont {Sbragaglia},\ and\ \citenamefont
  {Bernaschi}}]{pelusi2022tlbfind}%
  \BibitemOpen
  \bibfield  {author} {\bibinfo {author} {\bibfnamefont {F.}~\bibnamefont
  {Pelusi}}, \bibinfo {author} {\bibfnamefont {M.}~\bibnamefont {Lulli}},
  \bibinfo {author} {\bibfnamefont {M.}~\bibnamefont {Sbragaglia}},\ and\
  \bibinfo {author} {\bibfnamefont {M.}~\bibnamefont {Bernaschi}},\ }\bibfield
  {title} {\bibinfo {title} {Tlbfind: a thermal lattice boltzmann code for
  concentrated emulsions with finite-size droplets},\ }\href
  {https://doi.org/https://doi.org/10.1016/j.cpc.2021.108259} {\bibfield
  {journal} {\bibinfo  {journal} {Computer Physics Communications}\ }\textbf
  {\bibinfo {volume} {273}},\ \bibinfo {pages} {108259} (\bibinfo {year}
  {2022})}\BibitemShut {NoStop}%
\bibitem [{\citenamefont {Kawakatsu}\ and\ \citenamefont
  {Kawasaki}(1990)}]{kawakatsu_hybrid_1990}%
  \BibitemOpen
  \bibfield  {author} {\bibinfo {author} {\bibfnamefont {T.}~\bibnamefont
  {Kawakatsu}}\ and\ \bibinfo {author} {\bibfnamefont {K.}~\bibnamefont
  {Kawasaki}},\ }\bibfield  {title} {{\selectlanguage {en}\bibinfo {title}
  {Hybrid models for the dynamics of an immiscible binary mixture with
  surfactant molecules}},\ }\href
  {https://doi.org/10.1016/0378-4371(90)90287-3} {\bibfield  {journal}
  {\bibinfo  {journal} {Physica A: Statistical Mechanics and its Applications}\
  }\textbf {\bibinfo {volume} {167}},\ \bibinfo {pages} {690} (\bibinfo {year}
  {1990})}\BibitemShut {NoStop}%
\bibitem [{\citenamefont {Doi}(2011)}]{Doi_2011}%
  \BibitemOpen
  \bibfield  {author} {\bibinfo {author} {\bibfnamefont {M.}~\bibnamefont
  {Doi}},\ }\bibfield  {title} {\bibinfo {title} {Onsager’s variational
  principle in soft matter},\ }\href
  {https://doi.org/10.1088/0953-8984/23/28/284118} {\bibfield  {journal}
  {\bibinfo  {journal} {Journal of Physics: Condensed Matter}\ }\textbf
  {\bibinfo {volume} {23}},\ \bibinfo {pages} {284118} (\bibinfo {year}
  {2011})}\BibitemShut {NoStop}%
\bibitem [{\citenamefont {Doi}(2013)}]{doi2013soft}%
  \BibitemOpen
  \bibfield  {author} {\bibinfo {author} {\bibfnamefont {M.}~\bibnamefont
  {Doi}},\ }\href {https://books.google.co.uk/books?id=ccUaBj73PZsC} {\emph
  {\bibinfo {title} {Soft Matter Physics}}}\ (\bibinfo  {publisher} {OUP
  Oxford},\ \bibinfo {year} {2013})\BibitemShut {NoStop}%
\bibitem [{\citenamefont {Hardy}\ \emph {et~al.}(2024)\citenamefont {Hardy},
  \citenamefont {Daddi-Moussa-Ider},\ and\ \citenamefont
  {Tjhung}}]{hardy2024hybrid}%
  \BibitemOpen
  \bibfield  {author} {\bibinfo {author} {\bibfnamefont {A.~J.}\ \bibnamefont
  {Hardy}}, \bibinfo {author} {\bibfnamefont {A.}~\bibnamefont
  {Daddi-Moussa-Ider}},\ and\ \bibinfo {author} {\bibfnamefont
  {E.}~\bibnamefont {Tjhung}},\ }\bibfield  {title} {\bibinfo {title} {Hybrid
  particle-phase field model and renormalized surface tension in dilute
  suspensions of nanoparticles},\ }\href
  {https://doi.org/10.1103/PhysRevE.110.044606} {\bibfield  {journal} {\bibinfo
   {journal} {Phys. Rev. E}\ }\textbf {\bibinfo {volume} {110}},\ \bibinfo
  {pages} {044606} (\bibinfo {year} {2024})}\BibitemShut {NoStop}%
\bibitem [{\citenamefont {Doi}\ and\ \citenamefont
  {Edwards}(1986)}]{doi1986theory}%
  \BibitemOpen
  \bibfield  {author} {\bibinfo {author} {\bibfnamefont {M.}~\bibnamefont
  {Doi}}\ and\ \bibinfo {author} {\bibfnamefont {S.}~\bibnamefont {Edwards}},\
  }\href {https://books.google.co.uk/books?id=h-PvAAAAMAAJ} {\emph {\bibinfo
  {title} {The Theory of Polymer Dynamics}}},\ International series of
  monographs on physics\ (\bibinfo  {publisher} {Clarendon Press},\ \bibinfo
  {year} {1986})\BibitemShut {NoStop}%
\bibitem [{\citenamefont {Hoffmann}\ \emph {et~al.}(2009)\citenamefont
  {Hoffmann}, \citenamefont {Wagner}, \citenamefont {Harnau},\ and\
  \citenamefont {Wittemann}}]{Hoffmann2009}%
  \BibitemOpen
  \bibfield  {author} {\bibinfo {author} {\bibfnamefont {M.}~\bibnamefont
  {Hoffmann}}, \bibinfo {author} {\bibfnamefont {C.~S.}\ \bibnamefont
  {Wagner}}, \bibinfo {author} {\bibfnamefont {L.}~\bibnamefont {Harnau}},\
  and\ \bibinfo {author} {\bibfnamefont {A.}~\bibnamefont {Wittemann}},\
  }\bibfield  {title} {\bibinfo {title} {3d brownian diffusion of
  submicron-sized particle clusters},\ }\href
  {https://doi.org/10.1021/nn900902b} {\bibfield  {journal} {\bibinfo
  {journal} {ACS Nano}\ }\textbf {\bibinfo {volume} {3}},\ \bibinfo {pages}
  {3326} (\bibinfo {year} {2009})}\BibitemShut {NoStop}%
\bibitem [{\citenamefont {Kim}\ and\ \citenamefont
  {Karrila}(2005)}]{kim2005microhydrodynamics}%
  \BibitemOpen
  \bibfield  {author} {\bibinfo {author} {\bibfnamefont {S.}~\bibnamefont
  {Kim}}\ and\ \bibinfo {author} {\bibfnamefont {S.}~\bibnamefont {Karrila}},\
  }\href {https://books.google.co.uk/books?id=_8llnUUGo0wC} {\emph {\bibinfo
  {title} {Microhydrodynamics: Principles and Selected Applications}}},\
  Butterworth - Heinemann series in chemical engineering\ (\bibinfo
  {publisher} {Dover Publications},\ \bibinfo {year} {2005})\BibitemShut
  {NoStop}%
\bibitem [{\citenamefont {Jeffery}(1922)}]{jeffery1922motion}%
  \BibitemOpen
  \bibfield  {author} {\bibinfo {author} {\bibfnamefont {G.~B.}\ \bibnamefont
  {Jeffery}},\ }\bibfield  {title} {\bibinfo {title} {The motion of ellipsoidal
  particles immersed in a viscous fluid},\ }\href
  {https://doi.org/10.1098/rspa.1922.0078} {\bibfield  {journal} {\bibinfo
  {journal} {Proceedings of the Royal Society of London. Series A, Containing
  Papers of a Mathematical and Physical Character}\ }\textbf {\bibinfo {volume}
  {102}},\ \bibinfo {pages} {161} (\bibinfo {year} {1922})}\BibitemShut
  {NoStop}%
\bibitem [{\citenamefont {Lee}\ \emph {et~al.}(2014)\citenamefont {Lee},
  \citenamefont {Huh}, \citenamefont {Jeong}, \citenamefont {Shin},
  \citenamefont {Yun},\ and\ \citenamefont {Kim}}]{lee2014physical}%
  \BibitemOpen
  \bibfield  {author} {\bibinfo {author} {\bibfnamefont {D.}~\bibnamefont
  {Lee}}, \bibinfo {author} {\bibfnamefont {J.-Y.}\ \bibnamefont {Huh}},
  \bibinfo {author} {\bibfnamefont {D.}~\bibnamefont {Jeong}}, \bibinfo
  {author} {\bibfnamefont {J.}~\bibnamefont {Shin}}, \bibinfo {author}
  {\bibfnamefont {A.}~\bibnamefont {Yun}},\ and\ \bibinfo {author}
  {\bibfnamefont {J.}~\bibnamefont {Kim}},\ }\bibfield  {title} {\bibinfo
  {title} {Physical, mathematical, and numerical derivations of the
  cahn–hilliard equation},\ }\href
  {https://doi.org/https://doi.org/10.1016/j.commatsci.2013.08.027} {\bibfield
  {journal} {\bibinfo  {journal} {Computational Materials Science}\ }\textbf
  {\bibinfo {volume} {81}},\ \bibinfo {pages} {216} (\bibinfo {year}
  {2014})}\BibitemShut {NoStop}%
\bibitem [{\citenamefont {Cates}\ and\ \citenamefont
  {Tjhung}(2018)}]{cates2018theories}%
  \BibitemOpen
  \bibfield  {author} {\bibinfo {author} {\bibfnamefont {M.~E.}\ \bibnamefont
  {Cates}}\ and\ \bibinfo {author} {\bibfnamefont {E.}~\bibnamefont {Tjhung}},\
  }\bibfield  {title} {\bibinfo {title} {Theories of binary fluid mixtures:
  from phase-separation kinetics to active emulsions},\ }\href
  {https://doi.org/10.1017/jfm.2017.832} {\bibfield  {journal} {\bibinfo
  {journal} {Journal of Fluid Mechanics}\ }\textbf {\bibinfo {volume} {836}},\
  \bibinfo {pages} {P1} (\bibinfo {year} {2018})}\BibitemShut {NoStop}%
\bibitem [{\citenamefont {Markovich}\ \emph {et~al.}(2019)\citenamefont
  {Markovich}, \citenamefont {Tjhung},\ and\ \citenamefont
  {Cates}}]{Markovich_2019}%
  \BibitemOpen
  \bibfield  {author} {\bibinfo {author} {\bibfnamefont {T.}~\bibnamefont
  {Markovich}}, \bibinfo {author} {\bibfnamefont {E.}~\bibnamefont {Tjhung}},\
  and\ \bibinfo {author} {\bibfnamefont {M.~E.}\ \bibnamefont {Cates}},\
  }\bibfield  {title} {\bibinfo {title} {Chiral active matter: microscopic
  'torque dipoles' have more than one hydrodynamic description},\ }\href
  {https://doi.org/10.1088/1367-2630/ab54af} {\bibfield  {journal} {\bibinfo
  {journal} {New Journal of Physics}\ }\textbf {\bibinfo {volume} {21}},\
  \bibinfo {pages} {112001} (\bibinfo {year} {2019})}\BibitemShut {NoStop}%
\bibitem [{\citenamefont {Harris}\ \emph {et~al.}(2020)\citenamefont {Harris},
  \citenamefont {Millman}, \citenamefont {van~der Walt}, \citenamefont
  {Gommers}, \citenamefont {Virtanen}, \citenamefont {Cournapeau},
  \citenamefont {Wieser}, \citenamefont {Taylor}, \citenamefont {Berg},
  \citenamefont {Smith}, \citenamefont {Kern}, \citenamefont {Picus},
  \citenamefont {Hoyer}, \citenamefont {van Kerkwijk}, \citenamefont {Brett},
  \citenamefont {Haldane}, \citenamefont {del R{\'{i}}o}, \citenamefont
  {Wiebe}, \citenamefont {Peterson}, \citenamefont {G{\'{e}}rard-Marchant},
  \citenamefont {Sheppard}, \citenamefont {Reddy}, \citenamefont {Weckesser},
  \citenamefont {Abbasi}, \citenamefont {Gohlke},\ and\ \citenamefont
  {Oliphant}}]{harris2020array}%
  \BibitemOpen
  \bibfield  {author} {\bibinfo {author} {\bibfnamefont {C.~R.}\ \bibnamefont
  {Harris}}, \bibinfo {author} {\bibfnamefont {K.~J.}\ \bibnamefont {Millman}},
  \bibinfo {author} {\bibfnamefont {S.~J.}\ \bibnamefont {van~der Walt}},
  \bibinfo {author} {\bibfnamefont {R.}~\bibnamefont {Gommers}}, \bibinfo
  {author} {\bibfnamefont {P.}~\bibnamefont {Virtanen}}, \bibinfo {author}
  {\bibfnamefont {D.}~\bibnamefont {Cournapeau}}, \bibinfo {author}
  {\bibfnamefont {E.}~\bibnamefont {Wieser}}, \bibinfo {author} {\bibfnamefont
  {J.}~\bibnamefont {Taylor}}, \bibinfo {author} {\bibfnamefont
  {S.}~\bibnamefont {Berg}}, \bibinfo {author} {\bibfnamefont {N.~J.}\
  \bibnamefont {Smith}}, \bibinfo {author} {\bibfnamefont {R.}~\bibnamefont
  {Kern}}, \bibinfo {author} {\bibfnamefont {M.}~\bibnamefont {Picus}},
  \bibinfo {author} {\bibfnamefont {S.}~\bibnamefont {Hoyer}}, \bibinfo
  {author} {\bibfnamefont {M.~H.}\ \bibnamefont {van Kerkwijk}}, \bibinfo
  {author} {\bibfnamefont {M.}~\bibnamefont {Brett}}, \bibinfo {author}
  {\bibfnamefont {A.}~\bibnamefont {Haldane}}, \bibinfo {author} {\bibfnamefont
  {J.~F.}\ \bibnamefont {del R{\'{i}}o}}, \bibinfo {author} {\bibfnamefont
  {M.}~\bibnamefont {Wiebe}}, \bibinfo {author} {\bibfnamefont
  {P.}~\bibnamefont {Peterson}}, \bibinfo {author} {\bibfnamefont
  {P.}~\bibnamefont {G{\'{e}}rard-Marchant}}, \bibinfo {author} {\bibfnamefont
  {K.}~\bibnamefont {Sheppard}}, \bibinfo {author} {\bibfnamefont
  {T.}~\bibnamefont {Reddy}}, \bibinfo {author} {\bibfnamefont
  {W.}~\bibnamefont {Weckesser}}, \bibinfo {author} {\bibfnamefont
  {H.}~\bibnamefont {Abbasi}}, \bibinfo {author} {\bibfnamefont
  {C.}~\bibnamefont {Gohlke}},\ and\ \bibinfo {author} {\bibfnamefont {T.~E.}\
  \bibnamefont {Oliphant}},\ }\bibfield  {title} {\bibinfo {title} {Array
  programming with {NumPy}},\ }\href
  {https://doi.org/10.1038/s41586-020-2649-2} {\bibfield  {journal} {\bibinfo
  {journal} {Nature}\ }\textbf {\bibinfo {volume} {585}},\ \bibinfo {pages}
  {357} (\bibinfo {year} {2020})}\BibitemShut {NoStop}%
\bibitem [{cod()}]{code}%
  \BibitemOpen
  \href@noop {} {}\bibinfo {howpublished}
  {\url{https://github.com/AlexHardy0/Surfactants}}\BibitemShut {NoStop}%
\bibitem [{\citenamefont {van~der Sman}\ and\ \citenamefont {van~der
  Graaf}(2006)}]{van2006diffuse}%
  \BibitemOpen
  \bibfield  {author} {\bibinfo {author} {\bibfnamefont {R.~G.~M.}\
  \bibnamefont {van~der Sman}}\ and\ \bibinfo {author} {\bibfnamefont
  {S.}~\bibnamefont {van~der Graaf}},\ }\bibfield  {title} {\bibinfo {title}
  {Diffuse interface model of surfactant adsorption onto flat and droplet
  interfaces},\ }\href {https://doi.org/10.1007/s00397-005-0081-z} {\bibfield
  {journal} {\bibinfo  {journal} {Rheologica Acta}\ }\textbf {\bibinfo {volume}
  {46}},\ \bibinfo {pages} {3} (\bibinfo {year} {2006})}\BibitemShut {NoStop}%
\bibitem [{\citenamefont {Zong}\ \emph {et~al.}(2020)\citenamefont {Zong},
  \citenamefont {Zhang}, \citenamefont {Liang}, \citenamefont {Wang},\ and\
  \citenamefont {Xu}}]{zong2020modeling}%
  \BibitemOpen
  \bibfield  {author} {\bibinfo {author} {\bibfnamefont {Y.}~\bibnamefont
  {Zong}}, \bibinfo {author} {\bibfnamefont {C.}~\bibnamefont {Zhang}},
  \bibinfo {author} {\bibfnamefont {H.}~\bibnamefont {Liang}}, \bibinfo
  {author} {\bibfnamefont {L.}~\bibnamefont {Wang}},\ and\ \bibinfo {author}
  {\bibfnamefont {J.}~\bibnamefont {Xu}},\ }\bibfield  {title} {\bibinfo
  {title} {Modeling surfactant-laden droplet dynamics by lattice boltzmann
  method},\ }\href {https://doi.org/10.1063/5.0028554} {\bibfield  {journal}
  {\bibinfo  {journal} {Physics of Fluids}\ }\textbf {\bibinfo {volume} {32}},\
  \bibinfo {pages} {122105} (\bibinfo {year} {2020})}\BibitemShut {NoStop}%
\bibitem [{\citenamefont {Faria}\ and\ \citenamefont
  {Vishnyakov}(2022)}]{Vishnyakov_2022}%
  \BibitemOpen
  \bibfield  {author} {\bibinfo {author} {\bibfnamefont {B.~F.}\ \bibnamefont
  {Faria}}\ and\ \bibinfo {author} {\bibfnamefont {A.~M.}\ \bibnamefont
  {Vishnyakov}},\ }\bibfield  {title} {\bibinfo {title} {Simulation of
  surfactant adsorption at liquid–liquid interface: What we may expect from
  soft-core models?},\ }\href {https://doi.org/10.1063/5.0087363} {\bibfield
  {journal} {\bibinfo  {journal} {The Journal of Chemical Physics}\ }\textbf
  {\bibinfo {volume} {157}},\ \bibinfo {pages} {094706} (\bibinfo {year}
  {2022})}\BibitemShut {NoStop}%
\bibitem [{\citenamefont {Jaensson}\ \emph {et~al.}(2017)\citenamefont
  {Jaensson}, \citenamefont {Hulsen},\ and\ \citenamefont
  {Anderson}}]{Jaensson_2017}%
  \BibitemOpen
  \bibfield  {author} {\bibinfo {author} {\bibfnamefont {N.}~\bibnamefont
  {Jaensson}}, \bibinfo {author} {\bibfnamefont {M.}~\bibnamefont {Hulsen}},\
  and\ \bibinfo {author} {\bibfnamefont {P.}~\bibnamefont {Anderson}},\
  }\bibfield  {title} {\bibinfo {title} {On the use of a diffuse-interface
  model for the simulation of rigid particles in two-phase newtonian and
  viscoelastic fluids},\ }\href
  {https://doi.org/https://doi.org/10.1016/j.compfluid.2017.06.024} {\bibfield
  {journal} {\bibinfo  {journal} {Computers and Fluids}\ }\textbf {\bibinfo
  {volume} {156}},\ \bibinfo {pages} {81} (\bibinfo {year} {2017})}\BibitemShut
  {NoStop}%
\bibitem [{\citenamefont {Saintillan}\ and\ \citenamefont
  {Shelley}(2013)}]{saintillan2013active}%
  \BibitemOpen
  \bibfield  {author} {\bibinfo {author} {\bibfnamefont {D.}~\bibnamefont
  {Saintillan}}\ and\ \bibinfo {author} {\bibfnamefont {M.~J.}\ \bibnamefont
  {Shelley}},\ }\bibfield  {title} {\bibinfo {title} {Active suspensions and
  their nonlinear models},\ }\href
  {https://doi.org/https://doi.org/10.1016/j.crhy.2013.04.001} {\bibfield
  {journal} {\bibinfo  {journal} {Comptes Rendus Physique}\ }\textbf {\bibinfo
  {volume} {14}},\ \bibinfo {pages} {497} (\bibinfo {year} {2013})}\BibitemShut
  {NoStop}%
\bibitem [{\citenamefont {Kairaliyeva}\ \emph {et~al.}(2017)\citenamefont
  {Kairaliyeva}, \citenamefont {Aksenenko}, \citenamefont {Mucic},
  \citenamefont {Makievski}, \citenamefont {Fainerman},\ and\ \citenamefont
  {Miller}}]{kairaliyeva2017surface}%
  \BibitemOpen
  \bibfield  {author} {\bibinfo {author} {\bibfnamefont {T.}~\bibnamefont
  {Kairaliyeva}}, \bibinfo {author} {\bibfnamefont {E.~V.}\ \bibnamefont
  {Aksenenko}}, \bibinfo {author} {\bibfnamefont {N.}~\bibnamefont {Mucic}},
  \bibinfo {author} {\bibfnamefont {A.~V.}\ \bibnamefont {Makievski}}, \bibinfo
  {author} {\bibfnamefont {V.~B.}\ \bibnamefont {Fainerman}},\ and\ \bibinfo
  {author} {\bibfnamefont {R.}~\bibnamefont {Miller}},\ }\bibfield  {title}
  {\bibinfo {title} {Surface tension and adsorption studies by drop profile
  analysis tensiometry},\ }\href
  {https://doi.org/https://doi.org/10.1007/s11743-017-2016-y} {\bibfield
  {journal} {\bibinfo  {journal} {Journal of Surfactants and Detergents}\
  }\textbf {\bibinfo {volume} {20}},\ \bibinfo {pages} {1225} (\bibinfo {year}
  {2017})}\BibitemShut {NoStop}%
\bibitem [{\citenamefont {Kleinheins}\ \emph {et~al.}(2023)\citenamefont
  {Kleinheins}, \citenamefont {Shardt}, \citenamefont {El~Haber}, \citenamefont
  {Ferronato}, \citenamefont {Nozière}, \citenamefont {Peter},\ and\
  \citenamefont {Marcolli}}]{kleinheins2023surface}%
  \BibitemOpen
  \bibfield  {author} {\bibinfo {author} {\bibfnamefont {J.}~\bibnamefont
  {Kleinheins}}, \bibinfo {author} {\bibfnamefont {N.}~\bibnamefont {Shardt}},
  \bibinfo {author} {\bibfnamefont {M.}~\bibnamefont {El~Haber}}, \bibinfo
  {author} {\bibfnamefont {C.}~\bibnamefont {Ferronato}}, \bibinfo {author}
  {\bibfnamefont {B.}~\bibnamefont {Nozière}}, \bibinfo {author}
  {\bibfnamefont {T.}~\bibnamefont {Peter}},\ and\ \bibinfo {author}
  {\bibfnamefont {C.}~\bibnamefont {Marcolli}},\ }\bibfield  {title} {\bibinfo
  {title} {Surface tension models for binary aqueous solutions: a review and
  intercomparison},\ }\href {https://doi.org/10.1039/D3CP00322A} {\bibfield
  {journal} {\bibinfo  {journal} {Phys. Chem. Chem. Phys.}\ }\textbf {\bibinfo
  {volume} {25}},\ \bibinfo {pages} {11055} (\bibinfo {year}
  {2023})}\BibitemShut {NoStop}%
\end{thebibliography}%

\end{document}